\documentclass[12pt]{article}
\usepackage{jheppub}
\usepackage{ifpdf}

\usepackage{amssymb}
\usepackage{color}
\usepackage{graphicx,psfrag, subfigure}
\usepackage{amsmath}
\usepackage{hyperref}
\usepackage{bbold}

\usepackage[applemac]{inputenc}
\usepackage{dsfont}

\usepackage{soul}
\definecolor{dm}{cmyk}{.20, 0, .30, 0}
\sethlcolor{dm}

\numberwithin{equation}{section}

\def\be{\begin{equation}}
\def\ee{\end{equation}}
\def\bea{\begin{eqnarray}}
\def\eea{\end{eqnarray}}

\def\M{M_{Pl}}

\begin{document}

\begin{titlepage}

\setcounter{page}{1} \baselineskip=15.5pt \thispagestyle{empty}

\bigskip\
\begin{center}
{\Large \bf On Gaussian Random Supergravity}\\

\vskip 5pt
\vskip 15pt
\end{center}
\vspace{0.5cm}
\begin{center}
{
Thomas C. Bachlechner}
\end{center}

\vspace{0.05cm}

\begin{center}
\vskip 4pt
\textsl{Department of Physics, Cornell University,
Ithaca, NY 14853 USA}

\end{center} %\vfil
{\small  \noindent  \\[0.2cm]
\noindent
We study the distribution of metastable vacua and the likelihood of slow roll inflation in high dimensional random landscapes. We consider two examples of landscapes: a Gaussian random potential and an effective supergravity potential defined via a Gaussian random superpotential and a trivial K\"ahler potential. To examine these landscapes we introduce a random matrix model that describes the correlations between various derivatives and we propose an efficient algorithm that allows for a numerical study of high dimensional random fields. Using these novel tools, we find that the vast majority of metastable critical points in $N$ dimensional random supergravities are either approximately supersymmetric with $|F|\ll M_{\text{susy}}$ or supersymmetric. Such approximately supersymmetric points are dynamical attractors in the landscape and the probability that a randomly chosen critical point is metastable scales as $\log(P)\propto -N$. We argue that random supergravities lead to potentially interesting inflationary dynamics.}

\vspace{0.3cm}

\vspace{0.6cm}

\vfil
\begin{flushleft}
\small March 18, 2014%\today
\end{flushleft}
\end{titlepage}

\tableofcontents
\section{Introduction}\label{intro}

String theory is the leading candidate for a fundamental theory to describe the universe we observe. It is crucial that a successful UV theory allows for a solution that is consistent with both historical and current observations. In particular, the universe appears to have evolved to its current state via a period of accelerated expansion \cite{Guth:1980zm,Sato:1980yn,Linde:1981mu}. The low energy effective theory of string theory is supergravity. Therefore, it is natural to ask whether a generic four dimensional supergravity theory can account for a small positive vacuum energy and a period of cosmic inflation. 

The explicit construction of a representative ensemble of low energy theories directly from string theory is still in the distant future. In order to study a large, perhaps even representative, class of supergravity theories we divert to an alternate approach governed by universality. In particular, the idea of a potential landscape in high dimensional field space marked the beginning of the study of statistical properties in low energy effective theories that originate from some unknown UV physics \cite{Douglas:2004zu,Denef:2004cf,Denef:2007pq,Susskind:2003kw,Agarwal:2011wm}. Effective field theories arising from string theory typically involve $N\gg 1$ scalar fields that enter as moduli of the internal manifold. In such high dimensional field spaces one expects central limit behavior leading to low energy observables that are largely independent of the detailed UV physics. A number of works have taken advantage of universality in Wilsonian effective theories. Some examples are Refs.~\cite{Feng:2000if,Bousso:2000xa,Ashok:2003gk, Tegmark:2004qd,Douglas:2004zu,Denef:2004cf,Easther:2005zr,Agarwal:2011wm,Frazer:2011br,Marsh:2011aa,McAllister:2012am,Bachlechner:2012at,Pedro:2013nda,Marsh:2013qca,Tye:2008ef}, in which a varying degree of structure from the underlying UV theory was taken into account.

In this work we continue the quest to describe both the local and global properties of random four dimensional ${\cal N}=1$ supergravity theories with a large number $N$ of complex scalar fields. In the past, statistical properties of supergravity theories were primarily studied locally \cite{Ashok:2003gk,Douglas:2004zu,Denef:2004cf,Marsh:2011aa,Bachlechner:2012at,Chen:2011ac,Rummel:2013yta,Danielsson:2012by}. The investigation of properties beyond isolated points in random landscapes, such as inflationary trajectories, was limited to a discrete choice of the potential and a small number of active fields, which obscured the structure of the effective supergravity potential. In this work we take a step towards describing local and global properties of high dimensional random supergravities, both analytically and numerically. We consider two types of random landscapes: the first landscape consists of a Gaussian random potential that is divorced from any underlying supergravity theory, while the second landscape is what we call a Gaussian random supergravity. The Gaussian random supergravity we consider arises by considering a superpotential comprised of a Gaussian random field while restricting to flat field space. Our ultimate goal is to understand the distribution of metastable vacua and properties of inflationary trajectories in high dimensional random supergravities.

Before we continue let us pause to precisely define the types of questions one may be interested in when discussing the vacuum distribution of random landscapes. Bousso and Polchinski observed in Ref.~\cite{Bousso:2000xa} that the possibility to choose fluxes in the internal manifold leads to a vast ensemble of potential landscapes\footnote{See also Ref.~\cite{Feng:2000if} for a different approach to obtain a small quantized unit in the effective cosmological constant.}. Assuming flux quanta $N^i\in {\mathbb Z}$, where $i=1.\dots, K$ and some effective metric $g_{ij}$ on moduli space the landscape can be schematically written as \cite{Bousso:2000xa,Denef:2007pq}
\be
V_{\vec{N}}=V_0(\vec{\phi})+\sum_{i,j}g_{ij}(\vec{\phi})N^iN^j\, .
\ee
Assuming that each potential, corresponding to a unique choice of flux, has a minimum value at $\vec{\phi}_*$, it is easy to see that the number of vacua with vacuum energy less than $\Lambda_*$ is given by the number of flux lattice points within a sphere of radius $R^2=|V_0|+\Lambda_*$. Thus, the distribution of cosmological constants scales exponentially with $K$ \cite{Denef:2007pq}. By this logic, string theory is consistent with an exponentially large number of vacua that can in principle account for the observed fine tuning of the cosmological constant. The Bousso-Polchinski argument is a statement about the ensemble of landscapes consistent with string theory (different flux choices) while referring only to local properties (the assumed existence of one vacuum). Note however that this argument counted potentials and assumed the existence of one (metastable) vacuum at $\vec{\phi}_*$. A metastable vacuum is a critical point at which the Hessian matrix is positive definite. Therefore, a more complete analysis should consider the fluctuation probability of Hessian eigenvalues at critical points to compute the relevant probability
\be\label{ensemblep}
P_{\text{ensemble}}(\text{metastable c.p.})={\langle P(\text{metastable c.p.})\rangle\over \langle P(\text{c.p.})\rangle}\, ,
\ee
where $\langle\dots\rangle$ indicates the ensemble average. That is, $P_{\text{ensemble}}(\text{metastable c.p.})$ is the probability that a randomly chosen critical point from a randomly chosen landscape is metastable. This quantity is a local property of the ensemble as only one point for each landscape is considered and the global structure (the existence of nearby vacua) is irrelevant. The study of local properties is relevant to answer the question of whether the Bousso-Polchinski argument in principle can account for the vast fine tuning of the cosmological constant. However, such an approach does not yield any information about the vacuum distribution in a single realization of the landscape. Therefore, another important quantity is the abundance of metastable critical points for one particular flux choice. We can define
\be
P_{\text{flux}}(\text{metastable c.p.})={\langle \text{\# of metastable c.p.}\rangle\over \langle \text{\# of c.p.}\rangle}\, ,
\ee
where again $\langle\dots\rangle$ indicates the ensemble average but now all critical points within a single landscape (i.e. single choice of flux) are counted. Furthermore, for both definitions of the metastability probability we can impose specific constraints. In particular we will focus on three cases: $P^{\text{generic}}$ gives the probability that a generic critical point is metastable, $P^{\text{approx. SUSY}}$ gives the probability that a critical point in the regime of approximate supersymmetry is metastable and $P^{\text{susy}}$ gives the probability that a supersymmetric point is metastable. Having defined the meaning of metastability we point out that in this work we only consider ensemble probabilities of metastability, defined in Eq.~(\ref{ensemblep}). The methods introduced in this work yield powerful tools to study the global metastability properties for a single flux choice and it will be interesting to investigate those properties in future work.

We develop and apply two separate sets of tools: a local random matrix description for random potentials and a novel, efficient method for the simulation of high dimensional random fields. A key observation is that the various derivatives of random fields are correlated. This correlation strongly affects the statistical properties of the resulting landscape. In Ref.~\cite{Marsh:2011aa} it was observed that the probability for metastability at generic points in a random supergravity scales as
\be
\log[P^{\text{generic}}_{\text{ensemble}}(\text{metastable c.p.})]\propto-N^2\, ,
\ee
which led to the conclusion that a vanishingly small fraction of generic critical points are metastable vacua. However, if there exists some non-generic class of critical points that has a larger probability for metastability, this species may dominate the ensemble of metastable points. Indeed, in this work we find that due to a particular correlation between the potential and the Hessian matrix, the probability for metastability approaches unity for relatively low lying critical points. This correlation is described by an intuitive statement: minima are low, maxima are high and saddles are at generic positions. The matrix description we introduce yields statistical properties that remain valid at non-generic points and thus allows for a detailed study of the ensemble of metastable vacua. In order to understand consequences for inflationary physics we propose a method to simulate high dimensional Gaussian random fields. The tools presented in this work will enable an efficient study of high dimensional random landscapes, including landscapes with non-trivial field space geometries.

\begin{figure}
  \centering
  \includegraphics[width=.71\textwidth]{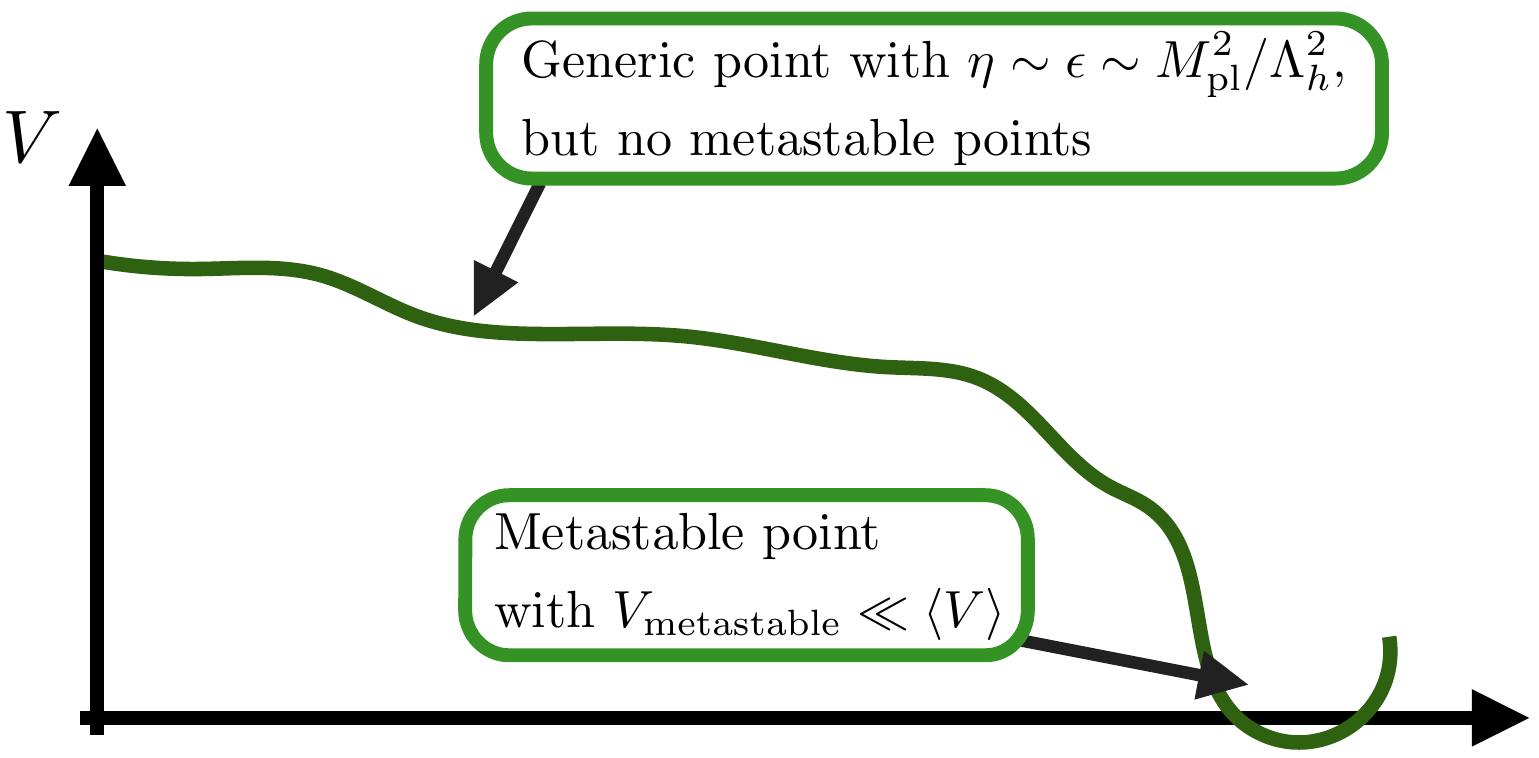}\label{fig:potscheme}
  \caption{\small Schematic depiction of the effective potential over a path starting at a generic point leading into a metastable minimum.}
\end{figure}

We find that the stability of critical points depends on the relative sizes of supersymmetry breaking and supersymmetric masses, in agreement with Ref.~\cite{Marsh:2011aa}. Furthermore, we find that at generic points where supersymmetry is badly broken metastability is unlikely. However, points of approximate supersymmetry are dynamical attractors where the probability for metastability is dramatically increased, yet still small. This provides an interesting mechanism for a decreasing vacuum energy as a metastable vacuum is approached. For the inflationary slow roll parameters we find $\langle\epsilon\rangle\sim \langle\eta\rangle \sim \M^2/\Lambda_h^2$, where $\Lambda_h$ is a horizontal scale in the superpotential. The landscape is schematically depicted in Figure \ref{fig:potscheme}.

The organization of this paper is as follows. In \S\ref{sec:landscaping} we begin by defining a Gaussian random landscape and a Gaussian random supergravity. These landscapes require a detailed understanding of the statistical properties of Gaussian random fields. Therefore, in \S\ref{sec:statGRF} we introduce a random matrix description of the various derivatives in Gaussian random fields and propose a novel mechanism for their numerical simulation. In \S\ref{rmtfgs} we apply these new tools to a simple ensemble of Gaussian random supergravities and study the distribution of metastable vacua. We discuss the possibility of slow roll inflation in the supergravity models in \S\ref{infl}. We conclude in \S\ref{conclusion}.

\section{Landscaping Effective Field Theories}\label{sec:landscaping}
In this section we will discuss two examples of random potentials that arise in effective field theories. In \S\ref{secGRL} we define an effective potential that is a Gaussian random field and briefly discuss some previous studies of similar landscapes. In \S\ref{sec:grsugra} we define a random landscape originating from four dimensional ${\mathcal{N}=1}$ supergravity with a Gaussian random superpotential.

Ultimately, we will be interested in statistical properties of the effective potential to study possible inflationary consequences and the distribution of metastable vacua. In order to simplify the study of Gaussian supergravities we discuss how the effective potential at generic points is related to a simple Gaussian random field.

\subsection{A Gaussian random landscape}\label{secGRL}

On general grounds, any potential for $N$ canonically normalized scalar fields can be written in the form
\be
V(\vec{\phi})=V_0+\Lambda_v^4f\left({\vec{\phi}\over \Lambda_h}\right)\,,
\ee
where $f$ is a dimensionless real function. In general, $f$ is not restricted to be of order one. However, in the absence of any additional known structure it is common to constrain $f$ to be of order one such that $\Lambda_v^4$ represents the vertical scale of a random potential, centered around some mean $V_0$ \cite{Tegmark:2004qd,Frazer:2011tg,Frazer:2011br,Marsh:2013qca}. It remains an open question as to what the expected mean and energy scales of a generic low energy effective potential are. In particular, it is not clear if $V_0$ scales with the number of fields. While in most of the literature $\langle V\rangle\ll\Lambda_v^4$ is assumed, this choice is far from obvious. Naively, one might expect $\langle V\rangle\sim \Lambda_v^4\lesssim \M^4$. However, in a Wilsonian effective quantum theory $V_0$ is a renormalized quantity that receives contributions from all masses in the theory. Therefore, if we consider a theory with $N$ species, it is not obvious that the expected value of $V_0$ is $N$ independent. For example, Dvali et al. argue in Refs.~\cite{Dvali:2007hz,Dvali:2012uq} that a theory with a large number of species at scale $\Lambda$ is technically {\it unnatural} unless
\be
\M^2\gtrsim N \Lambda^2\,.
\ee
In this work we leave $V_0$ as a free parameter that may depend on the number of fields $N$. This choice will become clear once we consider effective potentials arising from random supergravities. In these theories of local supersymmetry we will observe that the mean potential at a generic point scales with the number of fields: $V_0\propto N$. 

While the precise form of the potential is determined at high energies, it is essentially a random function at low energies. In the absence of any further information we are free to choose a landscape that is described by a stationary, isotropic Gaussian random field:
\bea\label{defgrf}
\langle V({\vec{\phi}}) \rangle&=&V_0\\
\langle (V({\vec{\phi}})-V_0)(V({\vec{\phi'}})-V_0) \rangle&=&c(|{\vec{\phi}}-{\vec{\phi}}'|)\,,
\eea
where $c(|{\vec{\phi}}-{\vec{\phi}}'|)$ is the covariance function, determining the correlations within the landscape. Although most results will generalize to more general cases we choose to consider a Gaussian covariance function in this work:
\be\label{covar2}
c(|{\vec{\phi}}-{\vec{\phi}}'|)=\Lambda_v^8 e^{-|{\vec{\phi}}-{\vec{\phi}}'|^2/\Lambda_h^2}\,.
\ee
Gaussian random fields are often expressed in terms of a superposition of Fourier modes \cite{Frazer:2011tg,Frazer:2011br,Marsh:2013qca}. We find such a representation impractical. To evaluate statistical properties analytically, the defining relations in Eq.~(\ref{defgrf}) are sufficient and easy to work with. More importantly, any decomposition in terms of Fourier modes on a lattice of size $L$ of dimension $N$ requires on the order of $L^N$ evaluations to obtain a numerical result. This is clearly impractical for the high dimensional potentials that we are interested in. Instead, in \S\ref{sec:numconstruction} we propose different tools to study high dimensional Gaussian random fields numerically, without referring to a Fourier decomposition on a lattice.

The choice of the landscape as a Gaussian random field with covariance (\ref{covar2}) leaves us with three free parameters that define the ensemble of potentials: the mean of the potential $V_0$, the horizontal scale $\Lambda_h$ and the vertical scale $\Lambda_v$. In this work we will explore the distribution of metastable vacua and consider the likelihood of inflation in Gaussian random landscapes, depending on the three scales. To implement such a study we will develop the required tools in \S\ref{sec:statGRF}.

\subsection{Defining a Gaussian random ${\mathcal N}=1$ supergravity}\label{sec:grsugra}

In the previous section we defined a landscape consisting of a Gaussian random field. In this section we discuss a landscape arising from the F-term potential of four dimensional ${\mathcal N}=1$ supergravity with $N$ complex scalar fields. 

The F-term potential is given by
\be
V=e^{K/\M^2}\left(F_a \bar{F}^a -{3\over\M^2}|W|^2\right)\,,
\ee
where $a=1,\dots, N$ labels the fields and $F_a=D_a W=(\partial_a+K_a/\M^2)W$. Derivatives of the K\"ahler potential are written as $\partial_a K=K_a$ and the K\"ahler metric is given by $\partial_a\partial_{\bar{b}} K=K_{a\bar{b}}$. Furthermore, we define the matrices of second and third derivatives as
\be
Z_{ab}\equiv {\mathcal D}_aF_b\,~~~\text{and}~~~U_{abc}\equiv {\mathcal D}_a{\mathcal D}_bF_c\, .
\ee
The F-term potential is fully defined in terms of the holomorphic superpotential and the K\"ahler potential, which we now address in turn.

While we are mostly agnostic about the UV physics that leads to the ensemble of effective supergravities, we now motivate the choice of superpotentials by considering ${\cal N}=1$ supersymmetric Calabi-Yau flux compactifications in type IIB string theory. The flux superpotential is linear in the flux and can be written as \cite{Gukov:1999ya,Ashok:2003gk}
\be
W(\phi)=\int_M \Omega\wedge G_3={\bf N}\cdot {\boldsymbol \Pi}(\phi)\,,
\ee
where ${\boldsymbol \Pi}$ are the periods of the holomorphic three form $\Pi_\alpha=\int_{\Sigma_\alpha} \Omega$ and ${\bf N}$ are the flux quanta. In explicit examples the periods ${\boldsymbol \Pi}$ can be computed. However, when considering a large number of contributions, the superpotential $W(\phi_a)$ is composed of a large number of essentially random terms and will obey central limit behavior, such that the distribution of $W$ can be approximated by a Gaussian random variable\footnote{Here we assume that the individual terms do not have heavy tails in their probability distributions such that the central limit theorem applies.}. Therefore, we propose to model the superpotential as a Gaussian random field defined by
\bea\label{superpotmodel}
\langle W(\phi)\rangle&=&W_0\,\nonumber\\
\langle (W(\phi)-W_0)({\overline{W}}(\bar{\phi}')-{\overline{W}}_0)\rangle&=&c(\phi,\phi')e^{K/\M^2}\, ,
\eea
where the first factor on the right hand side defines the correlation function and the second factor ensures the correct behavior of the superpotential under K\"ahler transformations\footnote{Ref.~\cite{Ashok:2003gk} suggests a natural ensemble of superpotentials of the form $\langle W(\phi)\overline{W}(\bar{\phi}')\rangle=e^{K(\phi,\bar{\phi})/\M^2}$. In this work we ignore the precise form of the K\"ahler potential and therefore we are free to choose a different ensemble of superpotentials.}. The model of the superpotential in Eq.~(\ref{superpotmodel}) deserves some discussion. First, the hope is to interpret the ensemble of superpotentials as effective data arising from UV physics. A stationary Gaussian random field is the appropriate description of a random process that at each point is described by a Gaussian random variable of constant mean. However, if the UV physics gives rise to heavy tails in the data the central limit theorem does not apply and the superpotential, despite being a large sum of random terms, will not converge to a Gaussian random variable. Furthermore, the correlation function $c(\phi,\phi')e^{K/\M^2}$ crucially defines the statistical properties of the superpotential ensemble. Therefore, the resulting low energy physics may depend heavily on the choice of the correlation function. It is beyond the scope of this work to determine the precise statistical properties of superpotentials arising from consisten string theory compactifications. Rather, we study a particular ensemble of superpotentials to study high dimensional random supergravity theories.

For small $\phi_a$, $\bar{\phi}_{\bar{a}}$ we can expand the K\"ahler potential around flat field space
\be
K(\phi_a,~\bar{\phi}_{\bar{a}})=\sum_{a=1}^N \phi_a \bar{\phi}^{a}+\sum_{n>2}{{\mathcal O}^n(\phi_a,~\bar{\phi}_{\bar{a}})\over \Lambda_K^{n-2}}\, ,
\ee
where ${\mathcal O}^n$ is an operator of dimension $n$ and $\Lambda_K$ is a mass scale. Thus, for $|\phi_a|\ll \Lambda_K$ the metric is just given by $K_{a\bar{b}}=\delta_{a\bar{b}}$. For simplicity we ignore non-trivial contributions to the K\"ahler potential and choose $\Lambda_K\rightarrow \infty$. This is a strong constraint on the models considered in this work. In typical flux compactifications the scale of higher order operators in the K\"ahler potential is small, $\Lambda_K\ll \M$ \cite{Giddings:2001yu}. While the study of more general K\"ahler potentials is interesting and will be the subject of a future work, here we constrain ourselves to a trivial K\"ahler potential
\be
K(\phi_a,~\bar{\phi}_{\bar{a}})= \sum_{a=1}^N \phi_a \bar{\phi}^{a}\,,  ~~~~~K_{a\bar{b}}=\delta_{a\bar{b}}\,,
\ee
with $|\phi|\lesssim\M$.

After fixing the K\"ahler gauge we now choose the two-point function of the superpotential to be
\be\label{eq:suppot}
\langle W(\phi)\rangle=W_0\,,~~~~~\langle (W(\phi)-W_0)(\overline{W}(\bar{\phi}')-\overline{W}_0)\rangle=\Lambda_v^6e^{-|\phi-\phi'|^2/\Lambda_h^2}\, ,
\ee
where $\Lambda_v$ is a mass scale determining the typical height of the superpotential, $\Lambda_h$ determines the horizontal scales and $W_0$ is the mean of the superpotential that is invariant under translations of $\phi$ and may be fixed by the UV physics. This choice of superpotential is equivalent to taking $W$ to be a Gaussian random holomorphic section with respect to the K\"ahler connection, as done in Ref.~\cite{Douglas:2004zu}. To have a well defined effective field theory we require $\Lambda_v,\Lambda_h\lesssim\M$. Here, we fixed the K\"ahler transformations, such that $K=0$ at the origin. With this the K\"ahler covariant derivative for $|\phi|\ll \M$ is given by
\be
D_aW=\left(\partial_a+{\bar\phi_a\over\M^2}\right)W\approx\partial_a W\, ,
\ee
and the effective potential simplifies to
\be\label{eq:effpot}
V(\phi,\bar\phi)\approx  |\partial_a W|^2-{3\over \M^2}|W|^2\,.
\ee
The first and second derivatives of the effective potential are given by \cite{Denef:2004cf}
\bea\label{eq:gradient}
\partial_a V&=&(\partial_a\partial_b W)\overline{\partial_b W}-{2\over \M^2}(\partial_a W)\overline{W}\\\label{eq:hessab}
\partial_a\partial_{b} V&=&(\partial_{abc}W)\overline{\partial_cW}-{1\over \M^2} (\partial_{ab}W)\overline{W}\\
\partial_a\partial_{\bar{b}} V&=&{\delta_{a\bar{b}}\over \M^2}\left(|\partial_aW|^2-{2\over \M^2}|W|^2\right)-{1\over\M^2}\partial_aW\overline{\partial_bW}+(\partial_a\partial^{\bar{c}}W)(\partial_{\bar{b}}\partial_{\bar{c}}\overline{W})\, .
\eea

Note that by choosing to model the superpotential as a Gaussian random field and limiting the discussion to a trivial K\"ahler metric, we are only left with three free parameters: $\Lambda_v$, $\Lambda_h$ and the mean of the superpotential $W_0$. In a metastable vacuum these three scales will set the supersymmetric masses and the scale of the supersymmetry-breaking soft masses. The supersymmetric masses, denoted by $M_{\text{susy}}$, are set by the scale of the eigenvalues of $Z\bar{Z}$, which generically is given by 
\be\label{msusy}
M_{\text{susy}}\sim \sqrt{N}{\Lambda_v^3\over \Lambda_h^2}\,.
\ee
At a metastable vacuum it is convenient to use the physical scale of supersymmetric masses, rather than the abstract quantity $\Lambda_h$. The supersymmetric mass scale is related to $\Lambda_h$ by
\be
\Lambda_h^2\equiv \sqrt{N}{\Lambda_v^3\over M_{\text{susy}}}\, .
\ee

In \S \ref{secGRL} we argued that in the absence of any underlying structure a generic landscape can be modeled as a Gaussian random field. We now imposed additional underlying structure, i.e. the supergravity effective potential, and following the reasoning of universality we should expect that at non-supersymmetric points we will recover a simple Gaussian random field description that breaks down as supersymmetric points are approached and the underlying structure becomes important. Indeed, in \S\ref{rmtfgs} we will find for the mean and variance of the random supergravity landscape at generic points
\bea
\langle V\rangle &=&2N{\Lambda_v^6\over \Lambda_h^2}
\\
\sigma_V&=&\sqrt{8N}{\Lambda_v^6\over \Lambda_h^2}
\, .
\eea
Furthermore, we will find $\sigma_{|\partial_aV|}=\sqrt{8 N}{\Lambda_v^6\over \Lambda_h^3}$. Therefore, we can approximate a random supergravity landscape in terms of a Gaussian random field $\tilde{V}$ as
\be
\langle \tilde{V}\rangle=\tilde{V}_0,~~\langle (\tilde{V}(\phi)-\tilde{V}_0)(\tilde{V}(\phi')-\tilde{V}_0)\rangle=\tilde{\Lambda}_v^4 e^{|\phi-\phi'|^2/\tilde{\Lambda}_h^2}\, ,
\ee
 where
 \be
\tilde{V}_0=2N{\Lambda_v^6\over \Lambda_h^2}\, ,~~~ \tilde{\Lambda}_v^4=\sqrt{8N} {\Lambda_v^6\over \Lambda_h^2}\, ,~~~\tilde{\Lambda}_h=\Lambda_h\, .
 \ee
In this manner, the landscape originating from supergravity can be viewed as a Gaussian random field in the non-supersymmetric limit. However, it is important to note that this approximation is only valid at generic points. Due to the various correlations in the supergravity landscape, the Gaussian random field approximation breaks down as a metastable vacuum is approached. We will find in \S\ref{rmtfgs} that at metastable vacua supersymmetry becomes important, which is consistent with the observation that the underlying structure of supergravity becomes relevant.

\section{Statistics of Gaussian Random Fields}\label{sec:statGRF}

In this section we will develop the tools required to investigate the statistical properties of Gaussian random fields both analytically and numerically. To study a Gaussian random field at a point, we develop a random matrix model that captures all correlations between derivatives of the field and allows for an analytic study in terms of random matrix ensembles (see e.g. Ref.~\cite{mehta2004random} for a pedagogical introduction to random matrix theory). While a random matrix model allows us to study Gaussian random fields at points, we are also interested in simulating high dimensional fields along trajectories. Therefore, we propose an efficient numerical algorithm to construct Gaussian random fields in high dimensional spaces.

We will find that at generic points the model of a GOE landscape introduced in Ref.~\cite{Marsh:2013qca} is a good approximation to the Hessian matrix of a Gaussian random field. However, away from generic points the Hessian matrix of a Gaussian random field exhibits correlations that dramatically change statistical observables\footnote{These correlations were observed before using a different approach in \cite{PhysRevLett.92.240601,PhysRevLett.109.167203,Bray:2007tf}.}. We will find that the vast majority of metastable critical points belongs to a species of non-generic points that have fluctuated to large or small values. Therefore the GOE landscape does not capture the vacuum statistics of Gaussian random fields.

\subsection{Random matrices in Gaussian random fields}\label{sec:analyticstatistics}
Suppose we have a stationary, isotropic and centered\footnote{This condition is easily relaxed by implementing a global, shift of the field.} random Gaussian field $V({\vec{\phi}})$ in $N$ dimensions. The statistical properties of the field are fully specified by
\bea
\langle V({\vec{\phi}}) \rangle&=&0\\\label{covariance}
\langle V({\vec{k}})V^*({\vec{ k}'}) \rangle&=&(2\pi)^{N} \delta^{N}({\vec{k}}+{\vec{k}}') P(k)\, ,
\eea
where we used the Fourier expansion of the field $V({\vec{\phi}})$:
\be\label{fourier}
V({\vec{\phi}})={1\over (2\pi)^{N}}\int d^N{\vec{k}} ~e^{i {\vec{k}}\cdot {\vec{\phi}}} V({\vec{k}})\,,
\ee
and $k=|{\vec{k}}|$. The two-point function in Eq.~(\ref{covariance}) can be rewritten as
\be\label{covspace}
\langle V({\vec{\phi}})V^*({\vec{\phi}}') \rangle=c(|{\vec{\phi}}-{\vec{\phi}}'|),~P({\vec{k}})=\int d^N{\vec{\phi}} ~e^{i {\vec{k}}\cdot {\vec{\phi}}}  c(|{\vec{\phi}}|)\, .
\ee
Using Eq.~(\ref{fourier}) we can now express the Hessian matrix $V_{ab}=\partial_{a}\partial_{b}V({\vec{\phi}})$ in terms of the Fourier components:
\be
V_{ab}=-{1\over (2\pi)^N}\int d^N{\vec{k}}~k_a k_b V({\vec{k}})e^{i {\vec{k}}\cdot {\vec{\phi}}}\, ,
\ee
which gives for the covariance tensor of the Hessian
\bea\label{covtensor}
\langle V_{ab}({\vec{\phi}})V^*_{cd}({\vec{\phi}})\rangle&=&{1\over (2\pi)^{N}}\int d^N{\vec{k}} ~k_{a}k_{b}k_{c}k_{d}P(k)\nonumber\\
&\propto&\delta_{ab}\delta_{cd}+\delta_{ad}\delta_{bc}+\delta_{ac}\delta_{bd}\, .
\eea

A plausible choice to model the Hessian matrix is the Wigner ensemble, i.e. the Hessian matrices are invariant under orthogonal transformations and the entries are independent and identically distributed random numbers \cite{Aazami:2005jf,Easther:2005zr,Marsh:2011aa,Bachlechner:2012at,Battefeld:2012qx,Pedro:2013nda,Marsh:2013qca}. The covariance tensor of the Wigner ensemble is given by
\be\label{covwig}
\langle H_{ab}({\vec{\phi}})H^*_{cd}({\vec{\phi}'})\rangle\propto \delta_{ad}\delta_{bc}+\delta_{ac}\delta_{bd}\,,
\ee
where $H$ is a Wigner matrix. Comparing Eq.~(\ref{covwig}) to Eq.~(\ref{covtensor}) we observe that the first term of the covariance tensor in a Gaussian random field is absent under the approximation that the Hessian matrix is in the Wigner ensemble. To understand this discrepancy, remember that the Wigner matrix was chosen under the assumption that the Hessian is independent of all other properties of the landscape. To relax this assumption let us consider the ensemble of Hessian matrices under the condition that the field $V$ takes on a particular value:
\be\label{constraint}
V({\vec{\phi}}_{0})=V_{0}\, .
\ee
Once the field is constrained to take on a particular value at $\vec{\phi}_0$, the eigenvalues of the Hessian are no longer drawn from the unbiased ensemble that is well approximated by a Wigner matrix with vanishing mean, but rather, by a new ensemble that is conditioned on our prior knowledge.

In order to evaluate expectation values for the ensemble under the constraint (\ref{constraint}) we need to rescale the field in order to satisfy Eq.~(\ref{covspace}) at $\vec{\phi}_0$
\be
\tilde{V}({\vec{\phi}})={\Lambda_v^4\over V_{0}}V({\vec{\phi}})\,,
\ee
such that
\be
\langle \tilde{V}({\vec{\phi}}_{0})\rangle_{V_0} =\Lambda_v^4,~\langle\tilde{V}({\vec{\phi}}) \tilde{V}({\vec{\phi}}_{0})\rangle_{V_0} =c(|{\vec{\phi}}-{\vec{\phi}}_{0}|)\, ,
\ee
where we denote the average of the ensemble that satisfies (\ref{constraint}) as $\langle \dots\rangle_{V_0} $. For the original field we immediately have the ensemble average
\be\label{fieldav}
\langle V({\vec{\phi}}) \rangle_{V_0} ={c(|{\vec{\phi}}-{\vec{\phi}}_{0}|)\over \Lambda_v^8}V_{0}\, ,
\ee
as expected.
Using Eq.~(\ref{fieldav}) and the definition of the field we readily find the ensemble average of the Hessian matrix at points conditioned to $V({\vec{\phi}}_{0})=V_{0}$:
\bea\label{hessshift}
\langle V_{ab}\rangle_{V_0} |_{{\vec{\phi}}={\vec{\phi}}_{0}}\nonumber &=&-{V_{0}\over \Lambda_v^4} {\delta_{ab}\over (2\pi)^{N}\Lambda_v^4}\int d^N{\vec{k}}~ k_{a}^{2}e^{i {\vec{k}}\cdot {\vec{\phi}}} P(|{\vec{k}}|) \\ 
&=&{V_{0}} {c''(0)\over \Lambda_v^8}{\delta_{ab}}\,.
\eea
This is a key result. Eq.~(\ref{hessshift}) indicates that the eigenvalue spectrum of the Hessian matrix in a Gaussian random field is directly correlated with the value of the field. This is a crucial result as this diagonal contribution dominates the probability for all eigenvalues to fluctuate to positivity. Using the same relations as above we find for the covariance tensor of the Hessian
\bea\label{hesscovariance}
\langle V_{ab}({\vec{\phi}})V^*_{cd}({\vec{\phi}})\rangle_{V_0} |_{{\vec{\phi}}={\vec{\phi}}_{0}}&=&\left({V^{2}_{0}\over \Lambda_v^8}\delta_{ab}\delta_{cd}+\delta_{ad}\delta_{bc}+\delta_{ac}\delta_{bd} \right) {c^{(4)}(0)\over 3}\\
&\propto&{V^{2}_{0}\over \Lambda_v^8}\delta_{ab}\delta_{cd}+\delta_{ad}\delta_{bc}+\delta_{ac}\delta_{bd}\, .
\eea
This result makes it more apparent under which condition the approximation that the Hessian matrix is indeed a Wigner matrix is applicable. Only at vanishing $V_0$ the Hessian is indeed precisely a Wigner matrix. At any other point the Hessian receives a diagonal contribution that reproduces the covariance tensor (\ref{covtensor}).

Repeating the same computation as above for the ensemble average of the gradient gives
\be\label{eq:gradsig}
\langle V_a({\vec{\phi}}) \rangle_{V_0} |_{{\vec{\phi}}={\vec{\phi}}_{0}}=0\, ,~~~\langle V_a({\vec{\phi}})V^*_b({\vec{\phi}}) \rangle_{V_0} |_{{\vec{\phi}}={\vec{\phi}}_{0}}=-c''(0)\delta_{ab}\, .
\ee
Note that the gradient $V_a$ is independent of both the zeroth and second derivative of the field. However, for the third derivative one finds a correlation with the gradient, where we now consider the ensemble where the gradient at $\vec{\phi}_0$ is given by $V^0_a$
\be\label{eq:gradU}
\langle V_{abc}\rangle_{V^0_a} ={c^{(4)}(0) \over 3c''(0)}(\delta_{ab}V^0_c+\delta_{ac}V^0_b+\delta_{cb}V^0_a)\, .
\ee
This again signals an important correlation within the potential. For the covariance tensor of third derivatives in the unconstrained ensemble we have
\be\label{eq:Ucov}
\langle V_{abc}V^*_{def}\rangle={c^{(6)}(0)\over 15}(\delta_{ab}\delta_{cd}\delta_{ef}+\text{perm.})\, .
\ee

\subsection{Distribution of vacua}
In the previous sections we derived some simple statistical properties for correlations between various derivatives of Gaussian random fields: the zeroth and second derivatives are correlated via Eq.~(\ref{hessshift}) and the first and third derivatives are correlated\footnote{Of course, there exist correlations between higher order derivatives that we are not interested in.} via Eq.~(\ref{eq:gradU}). The covariance tensor of the Hessian is given by Eq.~(\ref{hesscovariance}), where neglecting the first term is equivalent to taking the approximation that the Hessian is a Wigner matrix. 

Using the random matrix model described above we can estimate the probability of extrema in Gaussian random fields\footnote{See also Ref.~\cite{Bray:2007tf} for an equivalent approach using the Coulomb gas picture of random matrix theory.}. From Eq.~(\ref{hesscovariance}) we find that the Hessian is well described by a Wigner matrix with variance $\sigma^{2}=2 c^{(4)}(0)/3$, shifted by an amount $\lambda_{0}=V_{0}c''(0)/\Lambda_v^8$. The eigenvalue density of a shifted Wigner matrix is given by the famous Wigner semi-circle law:
\be
\rho(\lambda)={1\over \pi N \sigma^{2}}\sqrt{2N\sigma^{2}-(\lambda-\lambda_{0})^{2}}\,.
\ee
Thus, once the eigenvalue distribution is shifted far enough to positive values so that the eigenvalue spectrum has vanishing overlap with negative eigenvalues, nearly every critical point at that field value will be a minimum. The field value $V_{c}$ that satisfies this constraint is given by
\be
0=2N\sigma^{2}-\lambda_{0}^{2}={4 N c^{(4)}(0)\over3}-\left({V_{c}c''(0)\over \Lambda_v^8} \right)^{2}\, ,
\ee
or
\be\label{critfield}
V_{c}=-\sqrt{4 N  c^{(4)}(0)\over3}{\Lambda_v^8\over c''(0)}\, .
\ee
In Ref.~\cite{Bray:2007tf} the density of minima in high dimensional Gaussian random fields has been calculated and the critical field value below which nearly all critical points are minima agrees with Eq.~(\ref{critfield}).

Now that we have obtained a rough estimate for the scale at which nearly all critical points will be minima we can make an estimate of the typical distance to a minimum from a generic point. To make this estimate we assume Euclidean field space and assume that the covariance function decays over a typical length scale $\Lambda_h$, such that points separated by a distance much greater than $\Lambda_h$ will be uncorrelated. This allows for a rough estimate of the typical distance to a minimum. A volume ${\mathcal V}$ contains a number $N_{c}$ critical points\footnote{Note that we only keep the exponential scaling.}:
\be
N_{c}\sim {{\mathcal V}\over \Lambda_h^{N}}e^{-N}\, .
\ee
Any critical point with a field value $V\lesssim V_{c}$ will most likely be a minimum. The probability that the field at a random critical point is less than the critical field is given by
\be
P(V\lesssim V_{c})=\int_{-\infty}^{V_{c}} dV~{1\over \sqrt{2\pi \Lambda_v^8}} e^{-{V^{2}\over 2\Lambda_v^8}}\, .
\ee
Assuming a Gaussian covariance function we have for the critical field value
\be
V_{c}=-2\sqrt{ N}\Lambda_v^4\, .
\ee
This gives at large $N$
\be
P(V\lesssim V_{c})\approx {1\over \sqrt{8\pi N}} e^{-2N}\,.
\ee
Thus, the typical distance to a minimum is given by
\be
X_{c}\sim \left ({\sqrt{8\pi N}  \Lambda_h^N }e^{3N}\right)^{1/N}\rightarrow {\Lambda_h e^{3}}\, ,
\ee
where in the last step the limit $N\rightarrow \infty$ is taken. Thus, even though the probability that the Hessian fluctuates to positivity at generic points is extremely small, because of additional correlations the typical distance to a minimum is of the same order as the correlation length of the field. This finding agrees with Ref.~\cite{Bray:2007tf}, where a different approach has been used to estimate the average distance between minima.

\subsection{A numerical approach to high dimensional Gaussian random fields} \label{sec:numcheck}
So far, we have only kept track of local properties of Gaussian random fields: the random matrix approach allows us to evaluate ensemble averages of various properties of the landscape. We now turn to understanding how to efficiently probe global properties of random fields.

One direct way to generate a Gaussian random field is to pick a basis of functions on a discrete lattice of size $L$ (corresponding to some IR and UV cutoff of the truncated Fourier series as $L=\Lambda_{\text{UV}}/\Lambda_{\text{IR}}$) and consider a superposition with random weights. This approach has been chosen in a series of works, see Refs.~\cite{Tegmark:2004qd,Frazer:2011tg,Frazer:2011br}. While it allows to generate a globally defined potential, it is impractical to study $N\gg1$ dimensional fields: the total number of terms required scales as $L^N$. 

Marsh et al. proposed another, more efficient algorithm to generate random landscapes in Ref.~\cite{Marsh:2013qca}. In Ref.~\cite{Marsh:2013qca} a GOE landscape is defined by demanding that the Hessian matrix is in Wigner's Gaussian orthogonal ensemble and evolves over field space via Dyson Brownian motion \cite{mehta2004random,Dyson}. This approach specifies the Hessian matrix along an arbitrary path, while the field itself is obtained by successive quadratic approximations. As the field is only specified along a trajectory, this approach requires only a relatively small number of evaluations, allowing for the study of high dimensional potentials. While Dyson Brownian motion has obvious computational advantages, it is important to recall that it imposes a very special structure on the potential and in general the potential is not well defined. Considering self intersecting paths leads to an inconsistency as Dyson-Brownian motion gives different values of the field for the same point. Furthermore, the potential is poorly bounded, as can be seen from a simple estimate: the probability for an eigenvalue fluctuation to positivity scales as $P(\lambda_{\text{min}}>0)\sim e^{-N^2}$. Assuming a typical horizontal scale in the potential $\Lambda_h$, at a generic point the distance to the closest minimum scales as $d_{\text{minumum}}\sim \Lambda_h e^N$. This is radically different from the result for a Gaussian random field, where the closest minimum is within a distance $d_{\text{minumum}}\sim \Lambda_h$. This discrepancy was expected from \S\ref{sec:analyticstatistics} where we saw that the GOE ensemble does not capture statistics at extrema of Gaussian random potentials.

In the following, we propose a novel method for efficiently simulating high dimensional, globally well defined Gaussian random fields.

\subsubsection{Progressive construction of Gaussian random fields}\label{sec:numconstruction}

As a first step towards studying inflationary trajectories that potentially include many fields and terminate in a (meta) stable vacuum, we consider the special case of multi-field evolution in a random Gaussian landscape defined by an arbitrary power spectrum. 
Recall that a stationary, isotropic Gaussian random field is defined by\footnote{For simplicity we set the ensemble average of $V$ to zero. A non-zero but stationary average is trivially achieved by adding a constant to the field.}
\be\label{GRFprop}
\langle V({\vec{\phi}})\rangle=0,~~\langle V({\vec{\phi}})V^*({\vec{\phi}}')\rangle=c(|{\vec{\phi}}'-{\vec{\phi}}|)\,,
\ee
where the vertical scale is set by $\sqrt{c(0)}=\Lambda_v^4$. In order to numerically study the statistical properties of a GRF with an $N$ dimensional parameter space, where $N\gg 1$ it is impractical to generate an explicit ensemble of fields over a fixed lattice as for any $N\gtrsim4$ the number of points required for evaluation becomes very large. Instead, in the following we demonstrate how to efficiently evaluate a GRF at any arbitrary point.

A collection $\{V({\vec{\phi}}_{1}),V({\vec{\phi}}_{2}),\dots \}$ is called a stationary, isotropic Gaussian random field if the properties (\ref{GRFprop}) are satisfied. Thus, we can generate such a collection iteratively, for any arbitrary ${\vec{\phi}}_{i}$. Under the assumption of isotropy we can arbitrarily choose an initial point ${\vec{\phi}}_{1}$. As no other points are specified, $V({\vec{\phi}}_{1})$ is required to be a Gaussian variable satisfying
\be
\langle V({\vec{\phi}}_{1})\rangle=0,~~\langle V({\vec{\phi}}_{1})^{2}\rangle=\Lambda_v^8\,,
\ee
i.e. it has a density function
\be
\rho_V[x]={1\over \sqrt{2\pi \Lambda_v^8}} e^{-x^{2}/2\Lambda_v^8}\,.
\ee
We abbreviate this by writing $V({\vec{\phi}}_{1})\sim \Omega(0,\Lambda_v^8)$. To add a new point to the collection we use the following ansatz:
\be\label{ansatz}
V({\vec{\phi}}_{i+1})=\sum_{j=1}^{i}\phi_{j}V({\vec{\phi}}_{j})+ \Omega(0,\sqrt{\Phi})\, ,
\ee
where we introduced the $i+1$ unknown variables $\phi_{i}$ and $\Phi$. Assuming that the $i$ elements $V({\vec{\phi}}_{i})$ form a GRF, we find with Eq.~(\ref{GRFprop})
\be
\langle V({\vec{\phi}}_{i+1})\rangle=\langle\sum_{j=1}^{i} \phi_{j}V({\vec{\phi}}_{j})+\Omega(0,\sqrt{\Phi}) \rangle=\sum_{j=1}^{i}\phi_{j} \langle V({\vec{\phi}}_{j})\rangle +\langle \Omega(0,\sqrt{\Phi}) \rangle=0\,.
\ee
Furthermore, from the second constraint in Eq.~(\ref{GRFprop}) we have $i+1$ equations. For $k=1,\dots,i$ we have
\bea
\langle V({\vec{\phi}}_{i+1})V^*({\vec{\phi}}_{k})\rangle&=&\left\langle\left(\sum_{j=1}^{i} \phi_{j}V({\vec{\phi}}_{j})+ \Omega(0,\sqrt{\Phi})\right)V^*({\vec{\phi}}_{k})\right\rangle\nonumber\\
&=&\sum_{j=1}^{i}\phi_{j} c(|\phi_{j}-\phi_{k}|)\, .
\eea
Defining the matrix $C_{ij}=c(|\phi_{i}-\phi_{j}|)$ and the vector $C^{i+1}_{k}=c(|\phi_{i+1}-\phi_{k}|)$ we find
\be\label{lineq}
\phi_{j}=(C_{jk})^{-1}C^{i+1}_{k}\,,
\ee
the sum over $k$ is implicit. Thus, the parameters $\phi_{j}$ can be determined by solving a system of $i$ linear equations. The last parameter $\Phi$ is found by considering the equation
\bea
\langle V({\vec{\phi}}_{i+1})^{2}\rangle&=&\langle ( \phi_{j}V({\vec{\phi}}_{j})+ \Omega(0,\sqrt{\Phi}))^{2}\rangle\nonumber\\
&=&\Phi+\sum_{l=1}^{i}\phi_{l}C^{i+1}_{l},\nonumber\\
&=&\Lambda_v^8\,,
\eea
where we used $C_{nn}=\Lambda_v^8$ and Eq.~(\ref{lineq}). Thus, we have for the parameter $\Phi$:
\be\label{phieq}
\Phi=\Lambda_v^8-\sum_{l=1}^{i}\phi_{l}C^{i+1}_{l}\, .
\ee
Concluding, if we have the first $i$ elements satisfying the requirements for a Gaussian random fields, the $i+1$st element is given by Eq.~(\ref{ansatz}), where the parameters are the solutions of $i+1$ linear equations Eq.~(\ref{lineq}) and Eq.~(\ref{phieq}). This allows for an efficient iterative construction of a GRF in an arbitrary number of dimensions. Note that this approach applies for an arbitrary field space geometry when the metric is known.

\subsubsection{Numerical Study of Hessian statistics in Gaussian Random Fields}
Now that we have established an efficient method to simulate a Gaussian random field iteratively, avoiding the large computational cost in high dimensional spaces, we are in a position to compare the analytic results of \S\ref{sec:analyticstatistics} to direct simulations. The goal of this section is to confirm the result that the Hessian of a Gaussian random field is given by a Wigner ensemble that is shifted by an appropriate amount to satisfy Eq.~(\ref{hessshift}) and Eq.~(\ref{hesscovariance}).

\begin{figure}
  \centering
  \includegraphics[width=.5\textwidth]{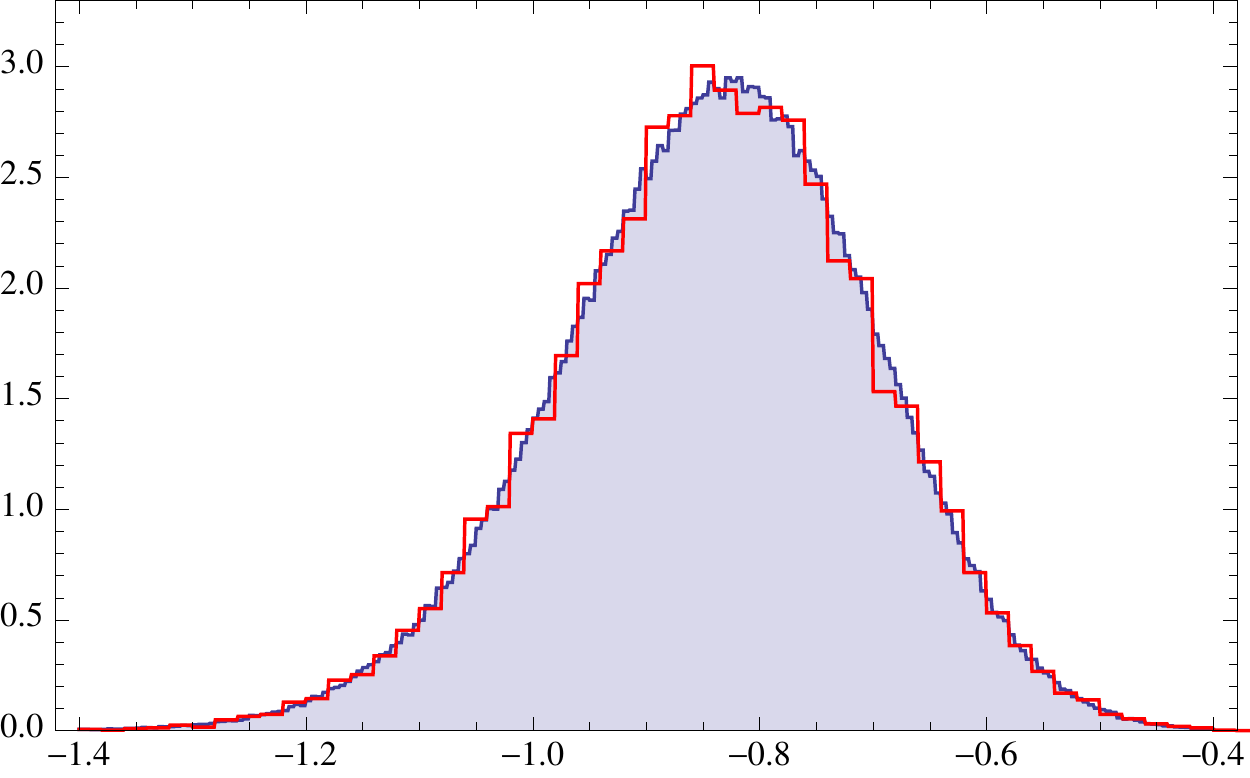}~~\includegraphics[width=.5\textwidth]{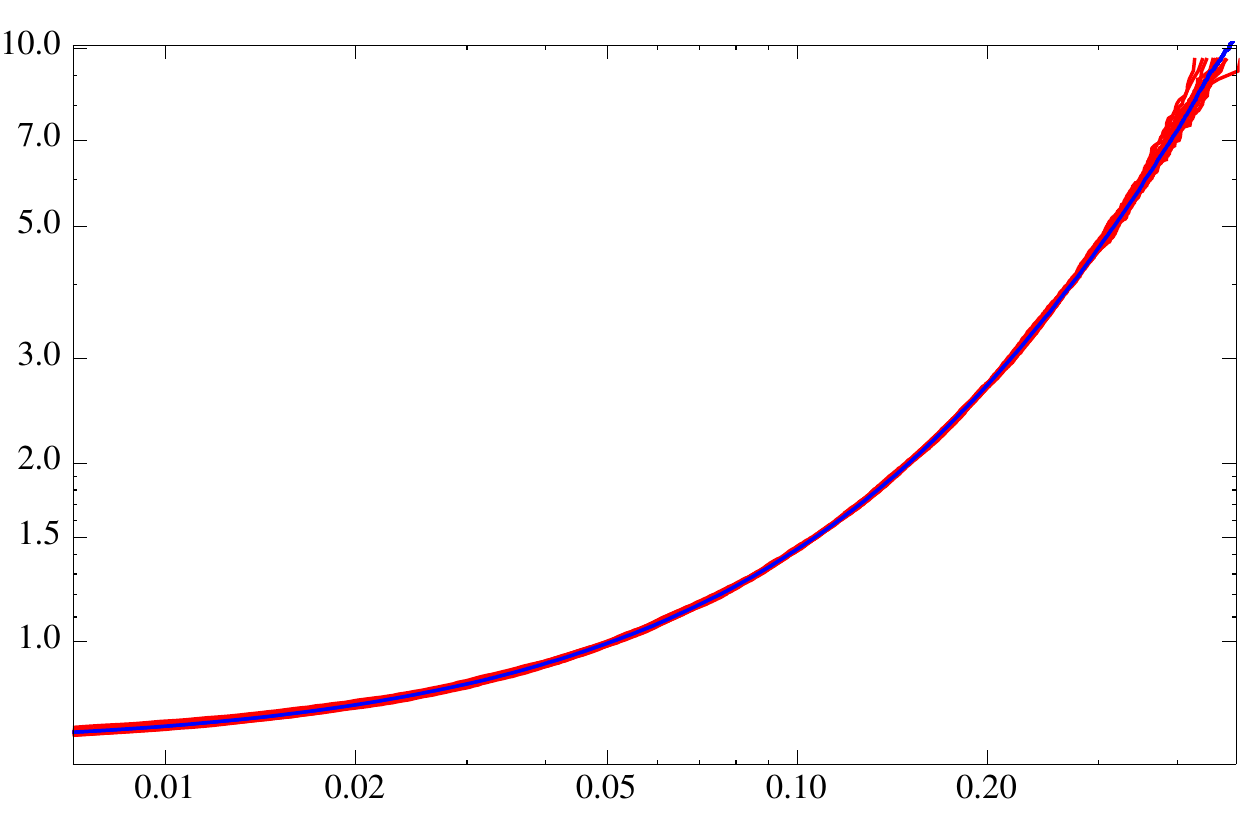}
  \caption{\small Left: Probability density function of the smallest Hessian eigenvalue for $N=10$, $V_{0}=8\Lambda_v^4$, shifted by the appropriate amount (red), along with numerical data from the Wigner ensemble (blue). Right: Negative logarithm of the fluctuation probability of the smallest Hessian eigenvalue to the right. All data is normalized by $\sqrt{2N\sigma^{2}}$.}\label{smallev}
\end{figure}

In the following we consider the specific Gaussian covariance function
\be
c(|{\vec{\phi}}|)=\Lambda_v^8 e^{-|{\vec{\phi}}|^{2}/\Lambda_h^2}\, ,
\ee
where $\Lambda_v^4$ sets the overall scale of the potential considered, while $\Lambda_h$ sets a horizontal scale. Let us consider a point ${\vec{\phi}}_{0}$ at which the potential is given by $V({\vec{\phi}}_{0})=V_{0}$. Using Eq.~(\ref{hessshift}) we expect for the mean of the entries of the Hessian matrix
\be
\langle V_{ab}\rangle_{V_0}|_{{\vec{\phi}}={\vec{\phi}}_{0}}=-2{V_{0}\over \Lambda_h^2}\delta_{ab}\, ,
\ee
and for the covariance tensor with Eq.~(\ref{hesscovariance})
\be
\langle V_{ab}({\vec{\phi}})V^*_{cd}({\vec{\phi}})\rangle_{V_0}|_{{\vec{\phi}}={\vec{\phi}}_{0}}=4 {\Lambda_v^8\over\Lambda_h^4}\left( {V^{2}_{0}\over \Lambda_v^8}\delta_{ab}\delta_{cd}+\delta_{ad}\delta_{bc}+\delta_{ac}\delta_{bd} \right)\, .
\ee
The probability density function of the Hessian eigenvalues $\lambda$ is given by
\be\label{eq:wigsp}
\rho(\lambda)={\Lambda_h^4\over 8 \pi N \Lambda_v^8}\sqrt{16 N{\Lambda_v^8\over \Lambda_h^4}-(\lambda+2{V_{0}\over \Lambda_h^2})^{2}}\,. 
\ee
\begin{figure}
  \centering
  \includegraphics[width=.7\textwidth]{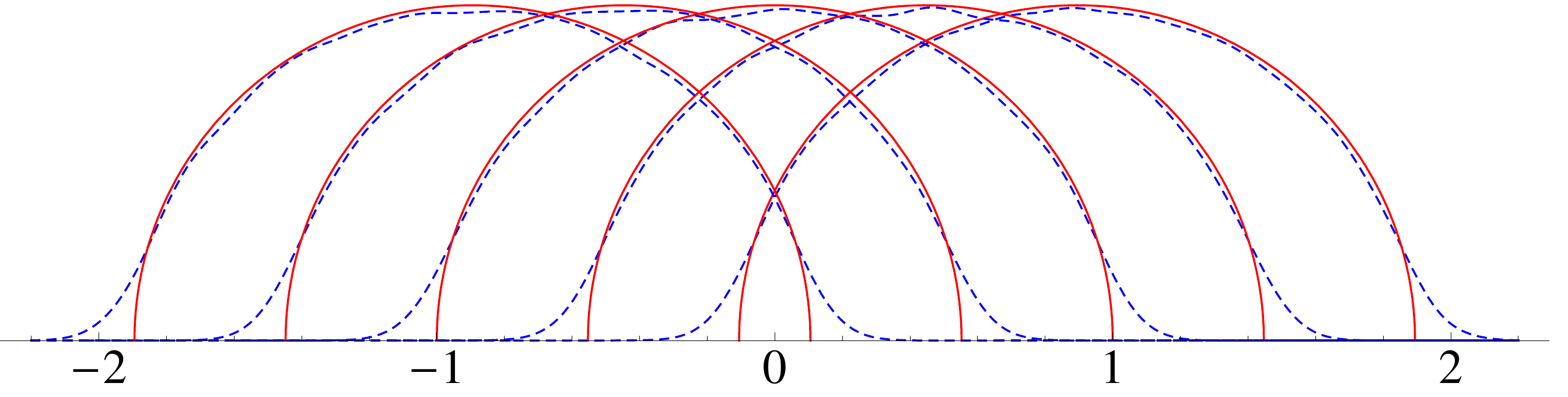}
  \caption{\small Hessian eigenvalue probability density function for a shifted Wigner ensemble from Eq.~(\ref{eq:wigsp}) (red) and $10^{4}$ numerically constructed Gaussian random landscapes with $N=20$ and $V_{0}/\Lambda_v^4=-8,-4,0,4,8$. The spectrum has been normalized by $\sqrt{2 N \sigma^{2}}$.}\label{fighesspdf}
\end{figure}
We compare the analytic probability density function for the Hessian eigenvalues to direct simulations of a Gaussian random field, assuming identical boundary conditions, in Figure \ref{fighesspdf}. The difference of the tail behavior is due to the fact that the Wigner semicircle law is only obtained in the large $N$ limit. While this limitation is present in the analytic expression for the semicircle law, the random matrix model still accurately describes a Gaussian random field, including small $N$ effects. To demonstrate this, the left part of Figure \ref{smallev} shows the probability density function of the smallest eigenvalue for the random matrix model and a direct simulation of a Gaussian random field ensemble. The right part of Figure \ref{smallev} shows the fluctuation probability of the smallest eigenvalue in both the random matrix model and the direct simulation.

To confirm Eq.~(\ref{hessshift}) we fit the mean of the eigenvalues of the Hessians to the model $\langle \lambda \rangle =-\mu{V_{0}\over \Lambda_h^2}$ and find numerically
\be
\mu=2.000\pm 3\times 10^{-3}\, .
\ee
It is clear from the data shown above that the random matrix model precisely matches the statistical properties of the Hessian matrix in Gaussian random fields. This was expected, as we constructed the random matrix model such that all correlation functions match. 

\begin{figure}
  \centering
  \includegraphics[width=.5\textwidth]{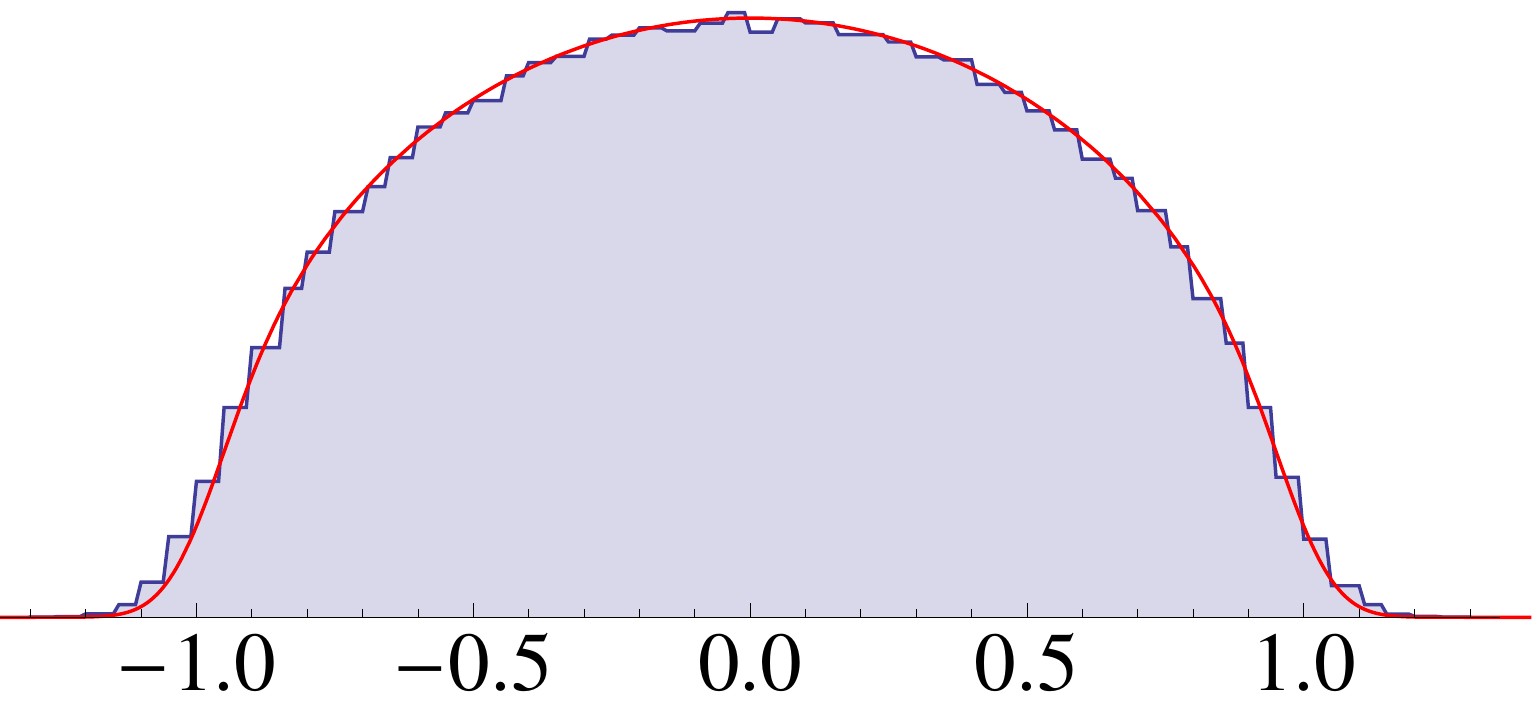}
  \caption{\small Probability distribution of the Hessian eigenvalues at random points in a GRF with $N=50$, along with the analytically obtained probability distribution in Eq.~(\ref{plambda}). All data is normalized.}\label{genpoint}
\end{figure}

So far we only evaluated the Hessian constrained to a particular value of the field. In order to obtain the distribution of the Hessian eigenvalues at a randomly chosen point we are required to evaluate the distribution of the variable $\lambda=\lambda_{\text{Wig}}+\lambda_{\text{shift}}$. This distribution is given by the convolution of the Wigner semicircle distribution with the Gaussian distribution determining the potential at a random point:
\bea\label{plambda}
\rho(\lambda)&=&\int d\mu~ \rho_{\text{Wigner}}(\mu)\rho_{\text{Gaussian}}(\lambda-\mu)\\
&=&{1\over 8 \pi N \Lambda_v^8/\Lambda_h^{4}} {1\over \sqrt{8\pi /\Lambda_h^{4} \Lambda_v^8}}\int_{-\infty}^{\infty} d\mu~\sqrt{16 N\Lambda_v^8/\Lambda_h^{4}-(\mu+2/\Lambda_h^{2}V_{0})^{2}}e^{-(\mu-\lambda)^{2}/(8/\Lambda_h^{4} \Lambda_v^8)}\nonumber\, .
\eea
Note that this expression clearly signals that the ensemble of Hessian matrices of a Gaussian random field at a random point is {\it not} given by a Wigner ensemble. In particular, the large fluctuation probability of Hessian eigenvalues scales as $e^{-N}$ and is dominated by the correlation to the field value. We numerically evaluate the integral in Eq.~(\ref{plambda}) and compare it to a simulation of random Gaussian fields in Figure \ref{genpoint}. 

\section{Random Matrix Theory for Gaussian Supergravity}\label{rmtfgs}

In this section we make use of the formalism introduced in \S\ref{sec:statGRF} to study the statistical properties of Gaussian random $\mathcal{N}=1$ supergravities, defined in \S\ref{sec:grsugra}. In \S\ref{sec:statgenpoint} and \S\ref{sec:numgenpoint} we discuss analytical and numerical results for the distribution of the potential and gradients, as well as a random matrix model for the Hessian matrix. In \S\ref{sec:statcritpoint} we impose the critical point condition and study the vacuum distribution and stability of critical points.

Recall that from Eq.~(\ref{eq:effpot}) the effective F-term potential in $\mathcal{N}=1$ supergravity under the assumption of a Gaussian random superpotential and trivial K\"ahler potential is given by

\be
V\approx  |F|^2-{3\over \M^2}|W|^2\, ,~~~~~~\text{for }~|\phi|\ll \M \, ,
\ee
where we defined $F_a=D_a W\approx \partial_aW=W_a$ and used K\"ahler transformations to set the potential to zero at $|\phi|\ll \M$.

\subsection{Statistics at non-critical points}\label{sec:statgenpoint}

As a first step, we obtain the probability distribution function of the potential at a random point. The statistical properties of the superpotential are given by Eq.~(\ref{eq:suppot}). As discussed in \S\ref{sec:analyticstatistics} the gradient of the superpotential is correlated with the matrix of third derivatives but is independent of both the value of $W$ and the Hessian matrix. Using Eq.~(\ref{eq:gradsig}) leads for $N\gg 1$ to the distributions
\bea\label{fdist}
\rho_{|F|^2}(x)&=&{1\over \sqrt{16\pi N}} {\Lambda_h^2\over \Lambda_v^6} \exp\left[ {-{(x-2{\Lambda_v^6\over \Lambda_h^2} N)^2\over 16N\Lambda_v^{12}/\Lambda_h^4}}\right]\,\\
\rho_{-3|W|^2/\M^2}(x)&=& {1\over \sqrt{-6\pi x\Lambda_v^6/\M^2}}e^{x/(6\Lambda_v^6/\M^2)}~~~\text{for $x<0$}\, ,
\eea
where we used the central limit theorem to approximate the chi-squared distribution of $|F_a|^2$ by a Gaussian distribution. It is clear that for large $N$ the second term in the potential is negligible, such that the probability distribution of the potential is given by
\be\label{rhov}
\rho_V(x)
\approx {1\over \sqrt{16\pi N}} {\Lambda_h^2\over \Lambda_v^6} \exp\left[ {-{(x-2{\Lambda_v^6\over \Lambda_h^2} N)^2\over 16N\Lambda_v^{12}/\Lambda_h^4}}\right]~~~~\text{for $N\gg \Lambda_h^2/\M^2$}\, .
\ee
Thus, we have for the ensemble average and standard deviation of the potential at generic points
\bea\label{avpot}
\langle V\rangle &=&2N{\Lambda_v^6\over \Lambda_h^2}
\\
\sigma_V&=&\sqrt{8N}{\Lambda_v^6\over \Lambda_h^2}
\, .
\eea
Note that for Gaussian random fields both the mean and the variance are independent of the field space dimension. Therefore, a Gaussian supergravity landscape is qualitatively different from a Gaussian random field.

To evaluate derivatives of the potential we require the statistical properties of the matrix $W_{ab}=\partial_a\partial_b W$. With the correlation function of the superpotential in Eq.~(\ref{eq:suppot}) and the covariance tensor in Eq.~(\ref{hesscovariance}) we have
\be
\langle W_{ab}\overline{W_{ab}}\rangle_W={4\Lambda_v^6\over \Lambda_h^4}(\delta_{ab}+1)+{4|W|^2\over \Lambda_h^4}\delta_{ab}\, .
\ee
We can model $W_{ab}$ by a complex symmetric matrix $Z_{ab}=\hat{Z}_{ab}\Lambda_v^3/\Lambda_h^2-2W/ \Lambda_h^2\mathbb{1}$ with independent entries of $\hat{Z}_{ab}$ distributed as
\be\label{zmatrix}
\hat{Z}_{ab}\in \Omega(0,\sqrt{4})~~~\text{for}~~~a\ne b~~~\text{and}~~~\hat{Z}_{aa}\in \Omega(0,\sqrt{8})~~~\text{(no sum on a)}\,.
\ee
The norm of the gradient is given by Eq.~(\ref{eq:gradient}). Assuming $N\gg 1$ and $|W|\ll \sqrt{N}\Lambda_v^3$ such that the shift of the Hessian due to large values of the superpotential is negligible, the probability densities of the individual terms are given by\footnote{Here the leading contribution to $W_{ab}$ comes from entries with standard deviation $\sqrt{4}\Lambda_v^3/\Lambda_h^2$ and vanishing mean.}
\bea\label{rhograd}
\rho_{\partial_a\partial_b W\overline{\partial_b W}}(x)&\approx&{1\over\sqrt{16 \pi N \Lambda_v^{12}/\Lambda_h^6}}e^{-{x^2\over 16 N \Lambda_v^{12}/\Lambda_h^6}}\\
\rho_{{2\over \M^2}(\partial_a W)\overline{W}}&\approx&{1\over\sqrt{16 \pi  \Lambda_v^{12}/(\Lambda_h^2\M^4)}}e^{-{x^2\over 16  \Lambda_v^{12}/(\Lambda_h^2\M^4)}}\, .
\eea
In the large $N$ limit the contribution from ${\partial_a\partial_b W\overline{\partial_b W}}$ in Eq.~(\ref{eq:gradient}) is dominant such that by using the asymptotic form of the chi distribution we have for the norm of the gradient
\be
\rho_{|\partial_aV|}\approx{1\over \sqrt{2\pi \sigma^2_{|\partial_aV|}}}\exp\left[{-{(x-\langle |\partial_aV|\rangle)^2\over 2 \sigma^2_{|\partial_aV|}}}\right]\, ,~~~\langle |\partial_aV|\rangle=\sqrt{8}N{\Lambda_v^6\over \Lambda_h^3}\,,~~~\sigma_{|\partial_aV|}=\sqrt{8 N}{\Lambda_v^6\over \Lambda_h^3}\, .
\ee
Finally, we are interested in a random matrix description of the Hessian matrix. Using Eq.~(\ref{eq:hessab}) we can write the Hessian as
\bea
{\mathcal H}&=&\left( \begin{array}{cc}\label{hessmat}
\partial^2_{a\bar{b}}V & \partial^2_{ab}V \\
\partial^2_{\bar{a}\bar{b}}V& \partial^2_{\bar{a}b}V \end{array} \right)\\
&=&\left( \begin{array}{cc}
Z_a^{~\bar{c}}\bar{Z}_{\bar{b}\bar{c}}-{1\over \M^2}F_a\bar{F}_{\bar{b}} ~~&~~ U_{abc}\bar{F}^{c}-{1\over \M^2}Z_{ab} \overline{W} \\
\bar{U}_{\bar{a}\bar{b}\bar{c}}F^{\bar{c}}-{1\over \M^2}\bar{Z}_{\bar{a}\bar{b}} W ~~& ~~\bar{Z}_{\bar{a}}^{~c}Z_{bc}-{1\over \M^2}\bar{F}_{\bar{a}}F_{b} \end{array} \right)+\nonumber \\&+&{\mathbb{1}\over \M^2}(|F|^2-{2\over \M^2} |W|^2)\, ,
\eea
where we defined $U_{abc}=\partial^3_{abc}W$. Recall that from Eq.~(\ref{eq:gradU}) and Eq.~(\ref{eq:Ucov}) the tensor of third derivatives of the superpotential $U_{abc}$ is correlated with $F_a$ and has a covariance tensor
\bea
\langle U_{abc}\rangle_{F_a} &=&-{2\over \Lambda_h^2}(\delta_{ab}F_c+\delta_{ac}F_b+\delta_{cb}F_a)\, \\
\langle U_{abc}\overline{U_{def}}\rangle &=&8{\Lambda_v^6\over \Lambda_h^6}(\delta_{ab}\delta_{cd}\delta_{ef}+\text{perm.})\, .
\eea
Thus, we can model $U_{abc}$ as a tensor
\be\label{umat}
U_{abc}=\hat{U}_{abc}\Lambda_v^3/\Lambda_h^3-{2\over \Lambda_h^2}(\delta_{ab}F_c+\delta_{ac}F_b+\delta_{cb}F_a)\, ,
\ee
where $\hat{U}_{abc}$ is a complex, totally symmetric tensor with entries distributed as
\bea
\hat{U}_{abc}\in& \Omega(0,\sqrt{8})~~&\text{for}~a\ne b\, ,~b\ne c\, ,~a\ne c\,\nonumber\\
\hat{U}_{aab}\in& \Omega(0,\sqrt{20})~~&\text{for}~a\ne b \,\nonumber\\
\hat{U}_{aaa}\in& \Omega(0,\sqrt{120})\, ,
\eea
where no sum is implied. While this non-trivial structure within the $U$ tensor makes it hard to study the spectrum of the Hessian analytically, we can consider the limit where $N\gg1$, such that the leading contributions in the Hessian are $Z\bar{Z}$, and $U\bar{F}$. The matrix $U\bar{F}$ has the following statistical properties
\bea
(U\bar{F})_{ab}&\in& \Omega(0,4\sqrt{N+5}{\Lambda_v^6\over\Lambda_h^4})~~~\text{for}~~~a\ne b\\
(U\bar{F})_{aa}&\in& \Omega(-4(N+2){\Lambda_v^6\over \Lambda_h^4},\sqrt{72N+456}{\Lambda_v^6\over\Lambda_h^4})\nonumber\, .
\eea
To obtain a rough estimate for the smallest eigenvalue of the Hessian, note that the matrix $Z\bar{Z}$ is positive definite. Ignoring the block diagonal component of the Hessian and approximating $(U\bar{F})_{ab}$ by a Wigner matrix we have for the left edge at fixed $W=W_0$ and\footnote{Note that by Eq.~(\ref{fdist}) the typical scale of $F$ is $|F|^2\sim 2N\Lambda_v^6/\Lambda_h^2$.} $|F|=|F_0|$
\be\label{minevw}
\lambda_{\text{min}}|_{W_0}= -\left( {4\over \Lambda_h^2}{\sqrt{1+{5\over N}}}-{1\over \M^2}\right)|F_0|^2+\left({4\over \Lambda_h^4}-{2\over \M^4}\right)|W_0|^2\, .
\ee

\subsection{Numerical results non-critical points}\label{sec:numgenpoint}

In the previous section we obtained some analytic results for statistical properties of the effective potential at generic points. We made a number of approximations. In particular, we used $N\gg1$ throughout, had to neglect the particular correlations of various quantities and were only able to make statements about generic points. However, it is interesting to see how the smallest eigenvalue of the Hessian may be correlated to other quantities, such as the value of the potential. These effects are difficult to obtain analytically. We now present a numerical study of the statistical properties of the potential at generic points. Here, we directly implement the Hessian matrix in Eq.~(\ref{hessmat}), including correlations between variables given in Eq.~(\ref{zmatrix}) and Eq.~(\ref{umat}). These results will be particularly interesting for the study of inflation, as they allow an estimate for the slow roll parameters in the potential considered. Note that it is computationally most efficient to implement the various quantities as random matrices even though we presented an algorithm to construct the potential iteratively in \S\ref{sec:numconstruction}. As we currently are concerned only with local statistical properties of the potential there is no reason to construct a global potential. We point out, however, that all results from matrix models are in excellent agreement with full simulations of the landscape.

\begin{figure}
  \centering
  \includegraphics[width=.5\textwidth]{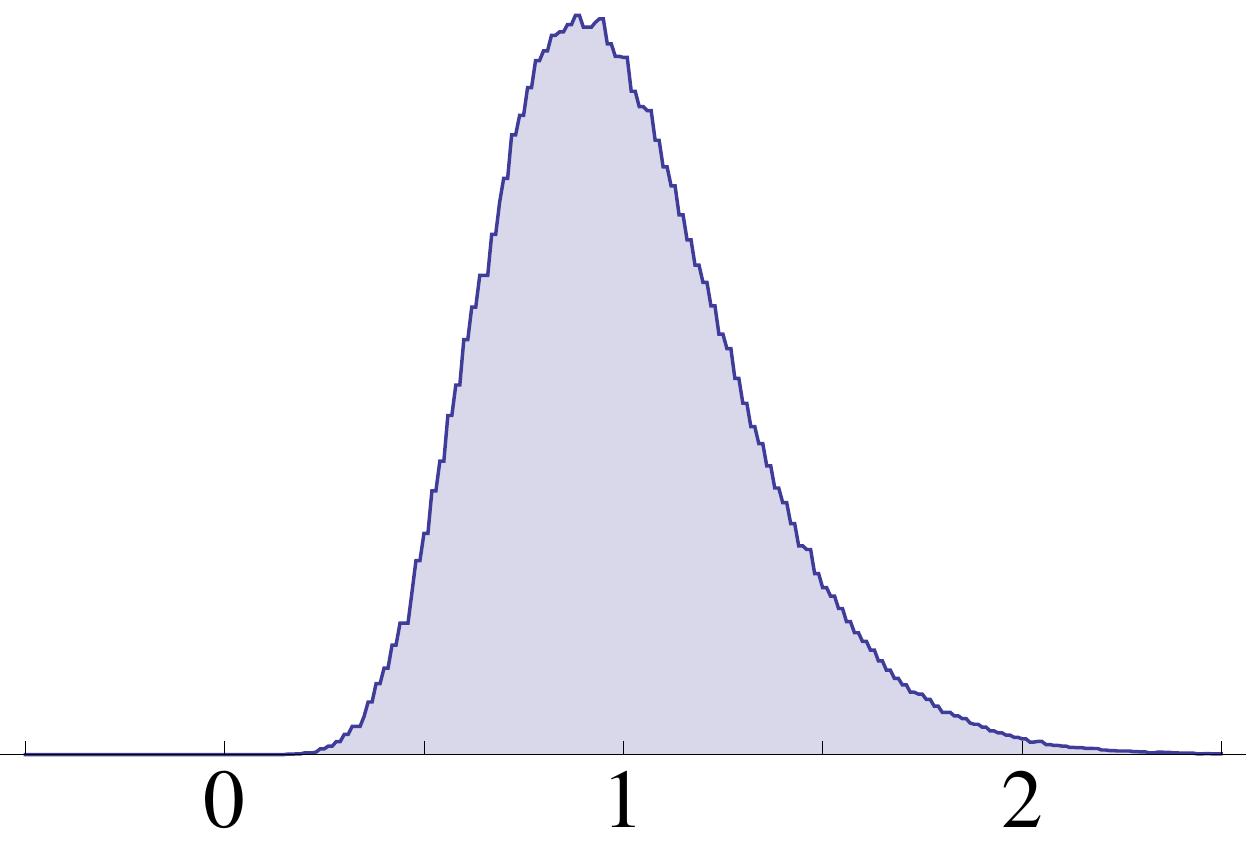}
  \caption{\small Distribution of the effective potential at generic points in units of $\langle V\rangle=2N\Lambda_v^6/\Lambda_h^2$ with $N=20$, see also Eq.~(\ref{rhov}).}\label{potdist}
\end{figure}
\begin{figure}
  \centering
  \includegraphics[width=.5\textwidth]{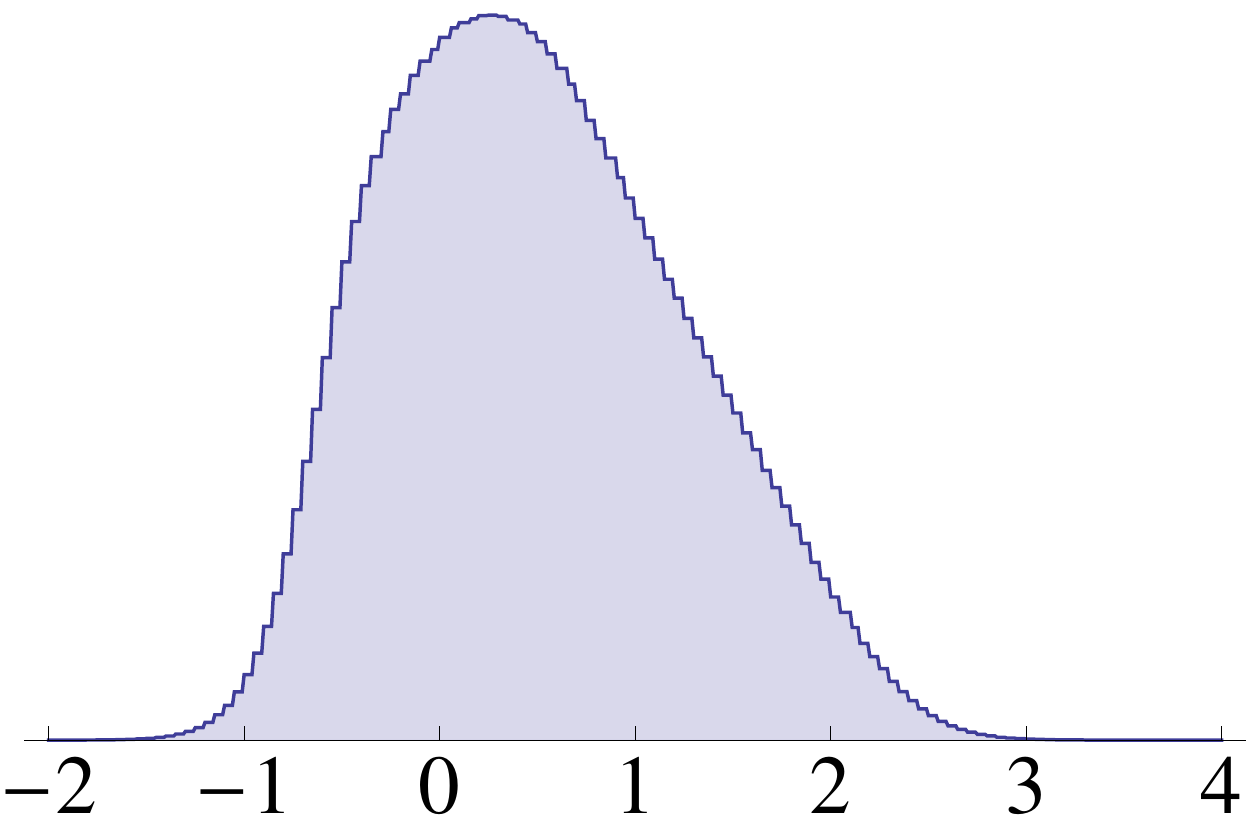}
  \caption{\small Spectrum of the eigenvalues of the full Hessian matrix ${\mathcal H}$ in Eq.~(\ref{hessmat}) for 
  $N=20$ in units of analytical result for the smallest eigenvalue, Eq.~(\ref{minevw}).}\label{hessspec}
\end{figure}
\begin{figure}
  \centering
  \includegraphics[width=.5\textwidth]{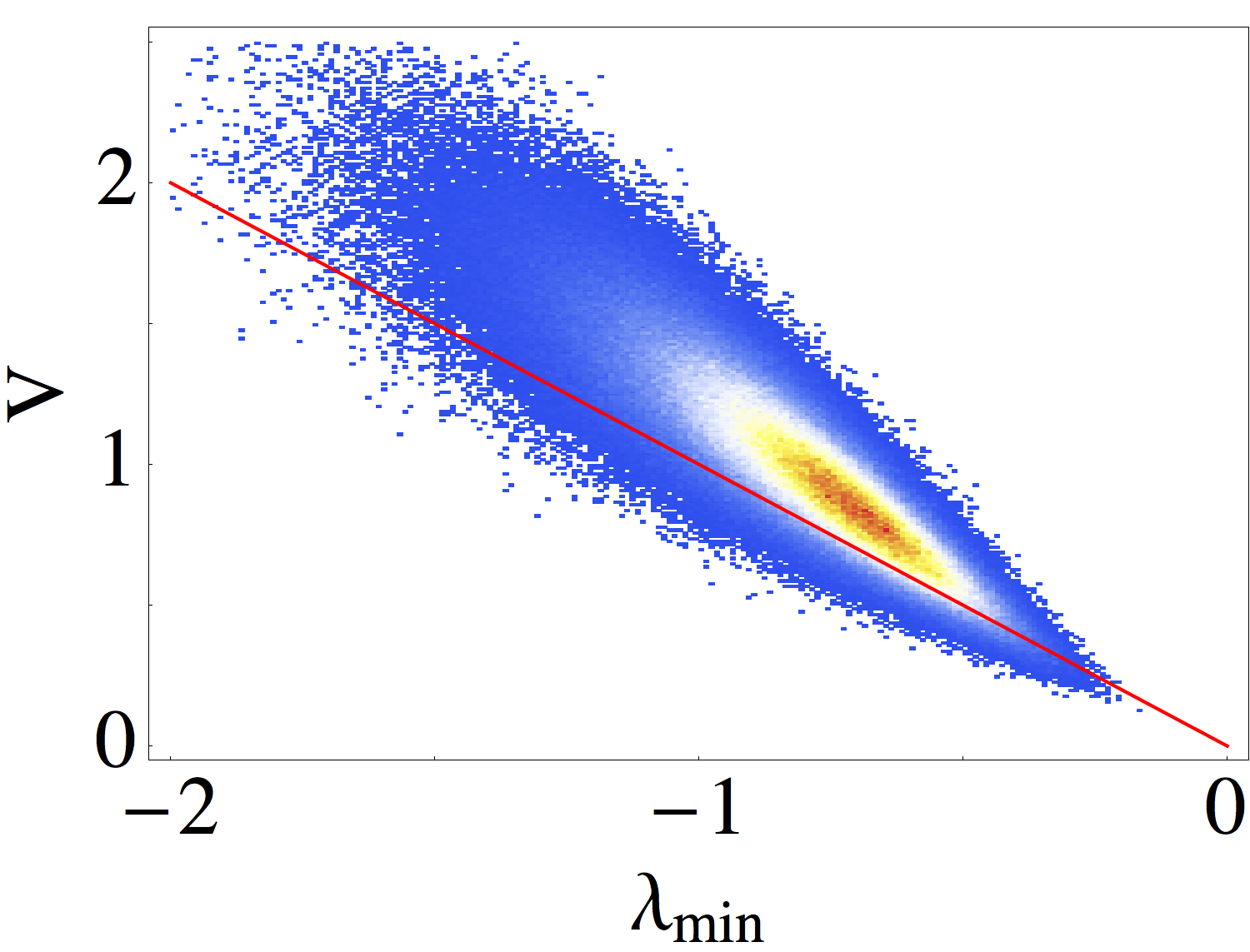}
  \caption{\small Smallest Hessian eigenvalue in units of $\lambda_{\text{min}}$ in Eq.~(\ref{minevw}) over the effective potential in units of $\langle V\rangle=2N \Lambda_v^6/\Lambda_h^2$ with $N=20$.}\label{minevpot}
\end{figure}

Figure \ref{potdist} shows the distribution of the effective potential in units of the average potential in Eq.~(\ref{avpot}). The numerical results are in excellent agreement with the analytical results for the mean and standard deviation of the potential.
Figure \ref{hessspec} shows the spectrum of the eigenvalues of the Hessian matrix at generic points in units of the analytical result for the smallest eigenvalue, Eq.~(\ref{minevw}). Note that Eq.~(\ref{minevw}) agrees within 2\% with the numerical result for the smallest eigenvalue.
Figure \ref{minevpot} shows the smallest eigenvalue of the Hessian over the value of the potential. The smallest eigenvalue is correlated with the potential, as expected. As observed for Gaussian random fields, the Hessian is shifted towards more positive eigenvalues with decreasing potential such that low lying critical points enjoy enhanced stability.


\subsection{Stability and distribution of critical points}\label{sec:statcritpoint}

In this subsection we discuss the distribution and stability of critical points in Gaussian random supergravities. It will turn out that the precise statistical properties at metastable critical points are hard to obtain. In this work we only present a first step towards studying realistic random supergravity theories. In particular, we neglect any contributions from non-trivial K\"ahler potentials, therefore we do not attempt any serious study of the vacuum distribution. Rather, we consider some heuristic arguments that hint towards an interesting vacuum distribution that warrants further study.

\begin{figure}
  \centering
  \includegraphics[width=.5\textwidth]{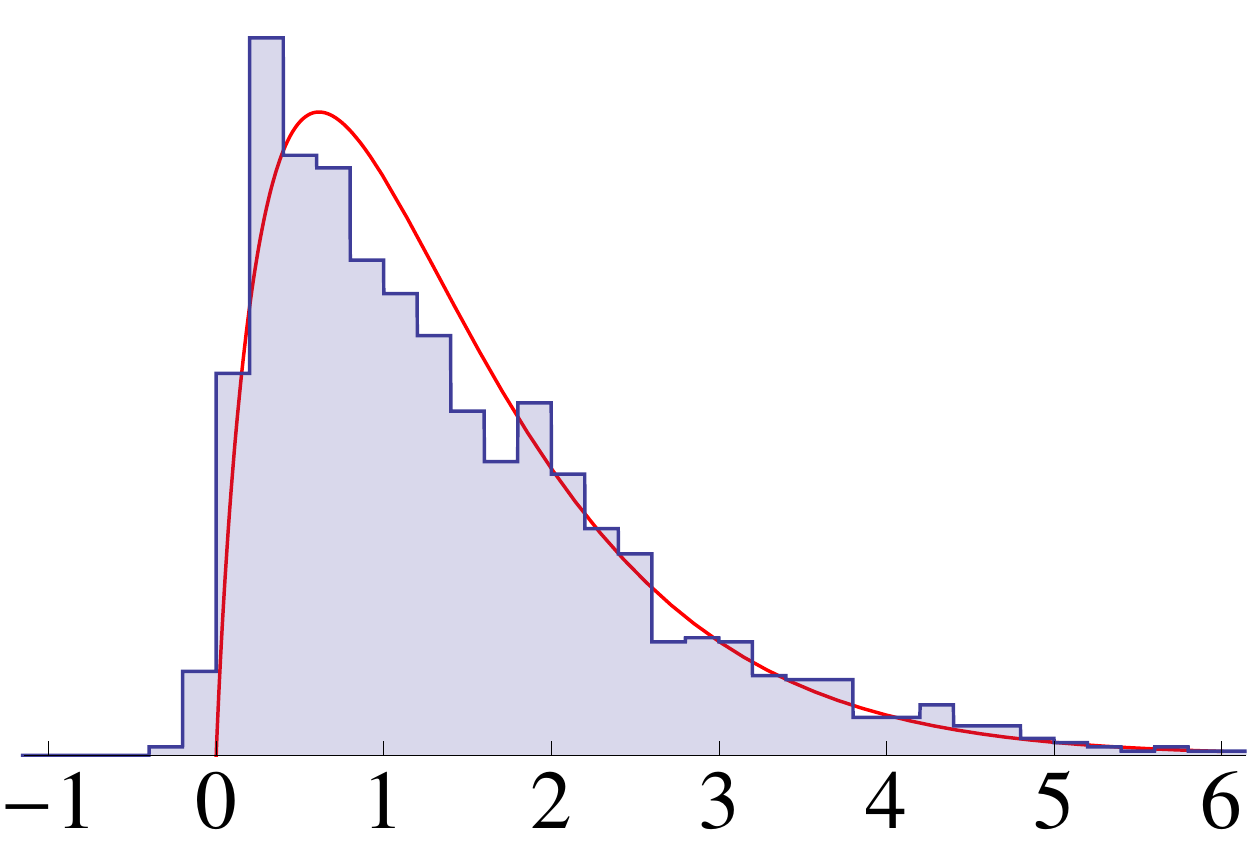}
  \caption{\small Distribution of the effective potential at generic metastable points in units of $\langle V_{\text{metastable}}\rangle$ in Eq.~(\ref{potmet}) $N=8$ along with the analytical estimate in Eq.~(\ref{metpdf}), evaluated including small $N$ effects.}\label{metcrit}
\end{figure}

With Eq.~(\ref{eq:gradient}) the critical point equation $\partial_a V=0$ can be written in matrix notation as
\be\label{cri}
Z\bar{F}={2\over \M^2}\overline{W}F\,.
\ee
Combining Eq.~(\ref{cri}) with its complex conjugate we have a condition on the eigenvectors and eigenvalues of $Z$:
\be
Z\bar{Z} F={4 |W|^2\over \M^4} F\, .
\ee
This imposes a constraint on the values of $W$ that are likely to be critical points: if $4|W|^2/\M^4$ is outside the support of the eigenvalue spectrum of $Z\bar{Z}$ then we have an additional suppression of the probability to find a metastable critical point, compared to that discussed in the previous subsection. Let us obtain an expression for the support of the matrix $Z\bar{Z}$. Recall that from Eq.~(\ref{zmatrix}) we can write
\be
Z=\hat{Z}\Lambda_v^3/\Lambda_h^2-2W/ \Lambda_h^2\mathbb{1}\, ,
\ee
where $\hat{Z}$ is a Wigner matrix. We now have
\be\label{zbarz}
Z\bar{Z} =\hat{Z}\hat{\bar{Z} }{\Lambda_v^6\over \Lambda_h^4}-2{\Lambda_v^3\over \Lambda_h^4}(W\hat{\bar{Z} }+\overline{W} \hat{Z})+4 {|W|^2\over \Lambda_h^4}\mathbb{1}\, .
\ee
To obtain an estimate for the support of the eigenvalue spectrum note that the first matrix is a Wishart matrix with $\sigma_{\hat{Z}\hat{\bar{Z} }}=\sqrt{4}\Lambda_v^3/\Lambda_h^2$. The eigenvalue spectrum of a Wishart matrix is given by
\be
\rho_{\hat{Z}\hat{\bar{Z} }}(\lambda)={1\over {2\pi N\sigma_{\hat{Z}\hat{\bar{Z} }}^2 \lambda}}\sqrt{(4 N \sigma_{\hat{Z}\hat{\bar{Z} }}^2-\lambda)\lambda}\,.
\ee
The second term in Eq.~(\ref{zbarz}) is a real Wigner matrix with $\sigma_{\text{Wig}}=2\sqrt{8}|W|\Lambda_v^3/\Lambda_h^4$ and eigenvalue spectrum
\be
\rho_{\text{Wig}}(\lambda)={1\over {2\pi N\sigma_{\text{Wig}}}} \sqrt{4N \sigma_{\text{Wig}}^2-\lambda^2}\,.
\ee
Combining the two spectra with the shift given by the last term in Eq.~(\ref{zbarz}) we have for the support the eigenvalue distribution of $Z\bar{Z}$
\bea\label{support}
&[8\Lambda_v^6/(\Lambda_h^4 N),16 N \Lambda_v^6/\Lambda_h^4]\,~~~\text{for }~|W|\ll \sqrt{2N}\Lambda_v^3\\
&[4|W|^2/\Lambda_h^4-8\sqrt{2N}|W|\Lambda_v^3/\Lambda_h^4,4|W|^2/\Lambda_h^4+8\sqrt{2N}|W|\Lambda_v^3/\Lambda_h^4]\,~~~\text{for }~|W|\gg \sqrt{2N}\Lambda_v^3\, .\nonumber
\eea
Outside the support of $Z\bar{Z}$ the probability to satisfy the critical point condition is exponentially suppressed. On the other hand, if $4|W|^2/\M^4$ is within the support of $Z\bar{Z}$, the critical point equation does not pose a significant constraint.
We now estimate the value of the superpotential at points that correspond to a positive definite Hessian matrix. The probability to satisfy the stability condition $\lambda_{min}>0$ can be written with Eq.~(\ref{minevw}) as
\be\label{metpdf}
P(\lambda_{min}>0)\propto P(\sigma_F|F|^2<x)P(\sigma_W|W|^2<x)\, , 
\ee
where 
\be
\sigma_F=\left({{4\over \Lambda_h^2}\sqrt{1+{5\over N}}} -{1\over \M^2}\right)\,,
\ee
and
\be
\sigma_W={4\over \Lambda_h^4}-{2\over \M^4}\, .
\ee
The parameter $x$ is chosen to maximize the fluctuation probability. Using Eq.~(\ref{fdist}) and taking the large $N$ limit we have
\be\label{psigf}
 P(\sigma_F|F|^2<x)\sim{x\Lambda_h^2\over 4\sqrt{N\pi}\Lambda_v^6\sigma_F}e^{-N/4}\, ,
\ee
and
\be\label{psigw}
P(\sigma_W|W|^2<x)\sim\sqrt{2\sigma_W\Lambda_v^6\over \pi x} e^{-x/(2\sigma_W \Lambda_v^6)}\, .
\ee
Using $|W|^2=x/\sigma_W$ and $|F|^2=x/\sigma_F$ we obtain an approximate expression for the probability distribution at metastable critical points
\be
\rho_{V_{\text{metastable}}}(x)\sim {\M^3 \Lambda_h^3 \sqrt{x} \over \sqrt{2\pi}\Lambda_v^9}\left({4\M^4-\Lambda_h^2\over 4\M^4-12\M^2\Lambda_h^2+\Lambda_h^4}\right)^{3/2}\exp\left[-{\M^2\Lambda_h^2(4\M^2-\Lambda_h^2)\over 2(4\M^4-12\M^2\Lambda_h^2+\Lambda_h^4)}x \right]\,.
\ee
From this probability distribution we obtain the ensemble average of the potential at metastable vacua
\bea\label{potmet}
\langle V_{\text{metastable}}\rangle&\sim&3{4\M^4-12\M^2\Lambda_h^2+\Lambda_h^4\over 4\M^4\Lambda_h^2-\M^2\Lambda_h^4}\Lambda_v^6\, .
\eea
At generic metastable vacua the superpotential takes on a generic value $|W|^2\sim\Lambda_v^6$. The above approximation of neglecting the critical point condition is consistent only when $4|W|^2/\M^4$ is within the support of the eigenvalue spectrum of $Z\bar{Z}$. Therefore, we obtain with Eq.~(\ref{support}) the consistency condition for the above analysis
\be
N\gg {4\over 3}{\M^4\over \Lambda_h^4}\,.
\ee
Assuming for now that $N\gg  {4\over 3}{\M^4\over \Lambda_h^4}$ we see from Eq.~(\ref{potmet}) that the mean potential at metastable critical points is independent of the number of fields. In particular, $\langle V_{\text{metastable}}\rangle\ll \langle V_{\text{generic}}\rangle$. This implies that metastable critical points occur at parametrically small values of the potential, while generic points in the potential will not be metastable. We are now in a position to compare this estimate to numerical simulations of the random matrix model for $N\gg  {4\over 3}{\M^4\over \Lambda_h^4}$. The results are shown in Figure \ref{metcrit}.

For the case where $N \lesssim{4\over 3}{\M^4\over \Lambda_h^4}$ we can only make qualitative statements. In this case the critical point equation imposes that one eigenvalue of $Z\bar{Z}$ is smaller than that of a typical Wishart matrix, introducing additional instability. Therefore, we expect the distribution of metastable critical points to peak at smaller values of the potential than estimated above. This finding is in qualitative agreement with the results of Ref.~\cite{Danielsson:2012by}.

\subsection{Approximate supersymmetry in Gaussian supergravity}
In the previous subsection we obtained an approximate distribution of metastable critical points in a Gaussian supergravity landscape. We found that for $N\gg  {4\over 3}{\M^4\over \Lambda_h^4}$ the majority of such metastable points has a positive effective potential and a fine tuned value of $|F|$. It was pointed out previously, that such approximately supersymmetric points lead to enhanced stability. Ref.~\cite{Marsh:2011aa} argued that while the probability of metastability at generic points scales approximately as $e^{-c N^2}$ it is conceivable that a tiny subclass of finely tuned critical points enjoys an enhanced likelihood of stability, such that this species would dominate the landscape of metastable vacua. Denef and Douglas found in Ref.~\cite{Denef:2004cf} that there exists an interesting species of critical points that are approximately supersymmetric, i.e. the F-terms are small compared to $M_\text{susy}$:
\be
\sqrt{3}{|W|}< F\ll {|Z_{ab}|}\sim{|U_{abc}|}\, .
\ee

We can compare this hierarchy to the expected values of the F-terms and the superpotential at metastable critical points from Eq.~(\ref{psigf}) and Eq.~(\ref{psigw}):
\be\label{fterm2}
\langle |F|\rangle_{\text{metastable}}\sim \sqrt{2\M^4-\Lambda_h^4\over 4 \M^2-\Lambda_h^2} {\sqrt{6}\Lambda_v^3\over \M\Lambda_h}\,,
\ee
and
\be
\langle |W|\rangle_{\text{metastable}}\sim\sqrt{3}\Lambda_v^3\,.
\ee
Comparing the scale of supersymmeic masses in Eq.~(\ref{msusy}) with Eq.~(\ref{fterm2}) implies exactly the hierarchy of approximate supersymmetry found by Denef and Douglas in Ref.~\cite{Denef:2004cf}: $|F|\ll M_{\text{susy}}$. Approximate supersymmetry enhances the likelihood of stability. Note that while $|F|$ is necessarily suppressed at a stable critical point, the above analysis only took into account the leading behavior around generic $|F|$ and, in particular, did not incorporate the requirement that the critical point equation be satisfied. Thus, we expect that the hierarchy found only gives a rough condition for metastable vacua. For example, AdS vacua with negative effective potentials (i.e. $|F|<\sqrt{3}|W|$) may constitute a large fraction of vacua for $N\lesssim {4\over 3}{\M^4\over \Lambda_h^4}$, when the analysis of \S\ref{sec:statcritpoint} becomes unreliable. The details are complicated and will be studied in future work.

\section{Towards Inflation in Random Landscapes}\label{infl}

In the previous sections we considered the spectrum of Hessian matrices in random landscapes. We now can consider a naive estimate for the possibility of inflation in high dimensional random landscapes. This question has been addressed in a series of previous works by assuming that the Hessian matrix is well approximated by a Wigner ensemble (see Refs.~\cite{Marsh:2013qca,Aazami:2005jf,Pedro:2013nda,Frazer:2011br,Battefeld:2013xwa}). This choice of Hessian ensemble made inflationary trajectories exponentially suppressed in the limit of a large number of scalar fields. In the following, we briefly review these arguments and consider the likelihood of inflation in Gaussian random landscapes.

We can roughly categorize the inflationary dynamics into two classes of models: large field models in which $|\Delta \phi|\sim \Lambda_h\gg \M$ and small field models with $|\Delta \phi|\sim \Lambda_h\ll \M$. In \S\ref{sec:landscaping} we found that the typical range of field evolution $|\Delta \phi|$ is not parametrically different from the typical length scale in the landscape. Therefore, we restrict the discussion to small field inflation models. This is a crucial difference compared to GOE landscapes considered in previous works \cite{Marsh:2013qca,Aazami:2005jf,Pedro:2013nda}. In GOE random potentials, the typical evolution is over a distance parametrically larger than $\Lambda_h$ until the fields settle into a minimum.

Let us start out with a flat FRW universe. Using local transformations to canonical kinetic terms we have the equations of motion for $N$ real scalar fields
\bea
\ddot{\phi}^a+3 H\dot{\phi}^a+V_a&=&0\, , \\
-{1\over 2 \M^2}\sum_{a}^{N}(\dot{\phi}^a)^2&=&\dot{H}\, ,\\
{1\over 2}\sum_{a}^{N}(\dot{\phi}^a)^2+V(\phi^a)&=&3 H^2\M^2\, .
\eea
To obtain a first approximation for inflationary background dynamics in the case of small field inflation where $|\Delta \phi|\ll \Lambda_h$ we only consider the quadratic expansion of the potential around a point in the landscape. For $\phi\lesssim \Lambda_h$, we can expand the potential as
\be
V(\phi^a)=\left(V_0+V_a{{\phi}^a}+{1\over 2} V_{ab}{{\phi}^{a}{\phi}^{b}}\right) \, ,
\ee
where $V_a=\partial_a V$. The equations of motions can be rewritten in terms of derivatives with respect to the number of e-folds $dN=Hdt$ 
\bea
{\phi}^{a''}+(3-\epsilon ){\phi}^{a'}+{1\over H^2}{\partial V\over\partial {\phi}^a}&=&0\, \\
{V\over \M^2 H^2}+{1\over 2\M^2}\sum_{a=1}^N ({\phi}^{a'})^2&=&3\, .
\eea
The slow roll parameters $\epsilon_V$ and $\eta_V$ are given for motion in the $a$ direction by
\be
\epsilon_V={\M^2\over 2}\left({V_a\over V}\right)^2\, ,~~\eta_V=\M^2{V_{aa}\over V}\,.
\ee

It would be very interesting to explore the full dynamics of inflationary trajectories in both Gaussian random landscapes and Gaussian random supergravities. While in principle in this work we presented all tools required for such a task, we delay the detailed study of full trajectories to a forthcoming work. Here, we merely introduce the tools and evaluate ensemble averages of the slow roll parameters, which will motivate a more detailed study of the classical and quantum evolution of the trajectory.

\subsection{Slow roll inflation in Gaussian random fields}
\begin{figure}
  \centering
  \includegraphics[width=.5\textwidth]{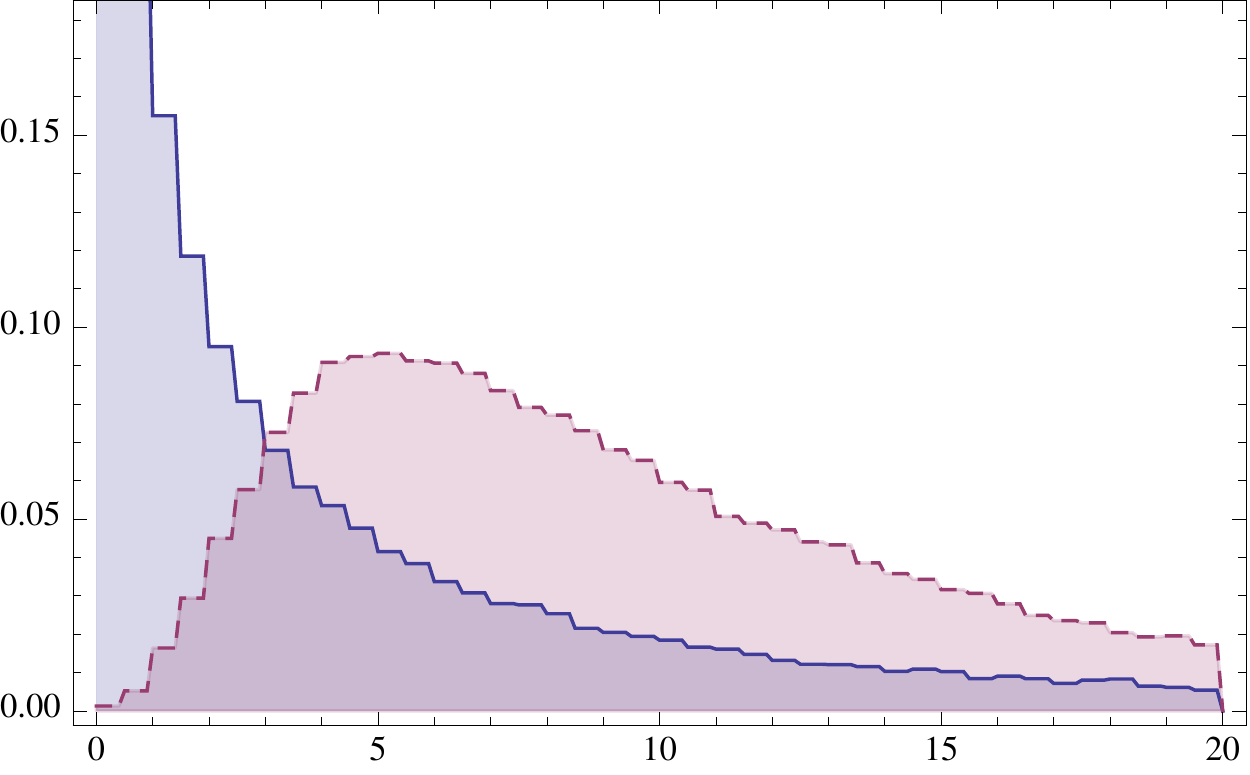}
  \caption{\small Probability density of typical slow roll parameters. Blue: $\epsilon$ Red, dashed: $\eta$ }\label{srparam1}
\end{figure}

In order to estimate the likelihood of slow roll inflation in a landscape modeled by a Gaussian random field we can consider a field with average $\bar{V}$:
\be
\langle V(\phi) \rangle=\bar{V}\, ,~~\langle (V(\phi)-\bar{V}) (V(\phi')-\bar{V})\rangle=\Lambda_v^8 e^{-|\phi-\phi'|^2/\Lambda_h^2}\, .
\ee
To estimate the slow roll parameters we expand the potential as
\be
V(\phi^a)=\Lambda_v^4\left(\hat{V}_0+{\hat{V}_a\over \Lambda_h}{{\phi}^a}+{1\over 2\Lambda_h^2} \hat{V}_{ab}{{\phi}^{a}{\phi}^{b}}\right) +\bar{V}\, .
\ee
Using the slow roll equations of motions we have for the slow roll parameters
\bea\label{epsilon}
\epsilon&\approx &{\M^2\over 2\Lambda_h^2}{|\hat{V}_a|^2\over (\hat{V}_0+\bar{V}/\Lambda_v^4)^2}\sim N{\M^2\over \Lambda_h^2(1+\bar{V}/\Lambda_v^4)^2}\, ,\\
\eta&\approx &{\M^2\over \Lambda_h^2}{(\text{Min}(\text{Eig}(\hat{V}_{ab}))-2\hat{V}_0)\over \hat{V}_0+\bar{V}/\Lambda_v^4}\sim -{\M^2\over \Lambda_h^2} {2(2\sqrt{N}-1)\over 1+\bar{V}/\Lambda_v^4} \, .\label{eta}
\eea
To simplify the discussion from now on, we only consider two cases for the shift of the potential. First, if $\bar{V}/ \Lambda_v^4\gtrsim 4\sqrt{N}\M^2/\Lambda_h^2$ the slow roll parameters are suppressed: $\epsilon,\eta\ll1$ as required for inflation. However, in this scenario the shift of the potential exceeds the critical potential height in Eq.~(\ref{critfield}) $V_c\sim \Lambda_v^4 2\sqrt{N}$, above which nearly all critical points will be extrema. Thus, any slow roll inflation occurring due to a high mean of the potential will terminate in eternal inflation with a large positive cosmological constant. On the other hand, choosing the potential to be centered around zero with $\bar{V}=0$ we can estimate how likely it is that the initial conditions for slow roll inflation are met. Following Ref.~\cite{Marsh:2013qca} there are two regimes in which the slow roll parameters are suppressed. Either inflation occurs by falling down a high slope, where the initial potential takes an unusually high value while the gradient and masses are of typical size, or inflation occurs at typical potential values while the gradient and masses fluctuate to allow for slow roll inflation. Note, however, that due to the additional shift of the smallest eigenvalue for high values of the potential in Eq.~(\ref{eta}), inflation down a high slope will never occur for typical masses and $\Lambda_h\lesssim \M$. This leaves fluctuations towards small gradients and masses as the only option for slow roll inflation. Assuming we require $\epsilon\lesssim \bar{\epsilon}$ and $\eta\lesssim \bar{\eta}$ with $\bar{V}=0$ and typical initial potential of $V_0\sim 1$ we have 
\bea
|\hat{V}_a|^2&\lesssim& {2\Lambda_h^2\over \M^2}\bar\epsilon\, ,\\
\text{Min}(\text{Eig}(\hat{V}_{ab}))&\lesssim &\bar\eta {\Lambda_h^2\over \M^2}-2\, .
\eea
The probability for this to occur is given by
\be
P\left(|\hat{V}_a|^2\lesssim {2\Lambda_h^2\over \M^2}\bar\epsilon \right)P\left(\text{Min}(\text{Eig}(\hat{V}_{ab}))\lesssim \bar\eta {\Lambda_h^2\over \M^2}-2 \right)\sim\left({4\Lambda_h^2\bar\epsilon\over N\pi \M^2} \right)^{N/2}Ae^{-\log(3)N^2/4}\, ,
\ee
where $A$ is a order one constant and we expanded the exponential assuming $-\bar\eta\Lambda_h^2/\M^2+2\hat{V}_0\ll \sqrt{N}$. Therefore, in a high dimensional Gaussian random landscape, inflationary points are extremely unlikely.
To illustrate the probability distribution of the slow roll parameters, Figure \ref{srparam1} shows the probability distribution for $\epsilon$ and $\eta$ for $\Lambda_h=\M$, $\bar{V}=0$ and $N=10$.

\subsection{Slow roll inflation in Gaussian random supergravity}
We now turn towards examining the possibility for slow roll inflation in Gaussian supergravity landscapes, defined in \S\ref{sec:grsugra}. As in the previous subsection, we can evaluate the slow roll parameters at generic points and find using Eq.~(\ref{rhov}) and Eq.~(\ref{rhograd}) 
\bea
\epsilon&=&{\M^2\over 2}{\langle |V_a|^2\rangle\over \langle V\rangle}\sim{\M^2\over \Lambda_h^2}\,\\
\eta&=& \M^2 {\langle\text{Min}(\text{Eig}(({\mathcal H}))\rangle\over \langle V\rangle}\sim 4{\M^2\over \Lambda_h^2}\,.
\eea
Comparing these slow roll parameters for a Gaussian supergravity to those of a Gaussian random field in Eq.~(\ref{epsilon}) shows that the slow roll parameters are smaller by a factor $N$ in the supergravity case. However, successful slow roll inflation requires $\eta\ll 1$. To examine if the likelihood of inflation is parametrically increased in the supergravity case we can consider the slow roll parameter at fixed $|F|$ and $|W|$ using Eq.~(\ref{minevw}) in the $N\gg 1$ limit
\be
\eta= \M^2 {\lambda_{\text{min}}\over V}\sim{|F|^2 \M^2 \Lambda_h^2 ( 4  \M^2-\Lambda_h^2 ) - 
 2 (\Lambda_h^4 - 2 \M^4) |W|^2\over \Lambda_h^4 (|F|^2 \M^2 - 3 |W|^2)}\, ,
\ee
which implies that for $\Lambda_h\ll \M$ the $\eta$ parameter at generic points is never is small enough to support a significant amount of inflation\footnote{While we only consider a centered Gaussian random superpotential, i.e. $\langle W\rangle=0$, it is easy to see from Eq.~(\ref{hessmat}) that a non-centered superpotential can only increase the $\eta$ parameter by shifting the smallest eigenvalue to lower values and decreasing the effective potential.}. In the estimate for the smallest eigenvalue we assumed non-fluctuated random matrices and generic points in the landscape. As the regime of approximate supersymmetry is approached the estimate for the smallest Hessian eigenvalue begins to break down as the subleading contributions to the Hessian become important. Therefore, the possibility for inflation at approximately supersymmetric points warrants further investigation. While $\Lambda_h\gtrsim \M$ allows for $\eta\lesssim1$ and inflation at generic points it is not clear that the supergravity approximation is valid in this regime.
 
We argued above that in the simple setup of a Gaussian random superpotential with trivial K\"ahler potential inflationary points are non-generic for $\Lambda_h\ll 1$. However, this may not be the final answer for more realistic random supergravities. The introduction of a non-trivial K\"ahler potential leads to additional contributions to all (K\"ahler covariant) derivatives and to the Hessian matrix \cite{Marsh:2011aa}. Furthermore, in this work we did not consider D-terms. A systematic study of these additional contributions is beyond the scope of this work and will be treated in a future project.

\section{Conclusion}\label{conclusion}
We studied the vacuum distribution and inflationary properties of high dimensional random landscapes. We considered landscapes consisting of a Gaussian random field and a toy four-dimensional ${\cal N}=1$ supergravity with $N\gg 1$ scalar fields and F-term supersymmetry breaking where the superpotential is a Gaussian random field and the K\"ahler potential is trivial. We constructed a random matrix model to study local properties of Gaussian random landscapes and proposed a novel algorithm that allows for an efficient numerical construction of high dimensional Gaussian random fields.

The various derivatives of a Gaussian random field are locally captured by correlated random matrix ensembles. In particular, we showed that the Hessian matrix is given by the Gaussian orthogonal Wigner ensemble with a diagonal contribution proportional to the value of the potential, while the tensor of third derivatives is correlated to the gradient of the potential. These correlations are crucial for the likelihood of a metastable vacuum: at a generic point the probability to encounter a metastable vacuum scales as $e^{-c N^2}$ for some order one constant $c$, while at points that are low in the potential the probability for metastability approaches unity. This comprises one of the crucial differences between Gaussian random fields and GOE landscapes introduced in Ref.~\cite{Marsh:2013qca}, where the Hessian is chosen to be an uncorrelated Wigner matrix. In the GOE landscape the distance to the closest minimum at a generic point scales as $e^{c N}\Lambda_h$, where $\Lambda_h$ is a horizontal length scale, while in Gaussian random fields, the distance to the closest minimum is roughly $\Lambda_h$. Therefore, GOE landscapes do not describe the approach to a minimum of a bounded random landscape. In this work we introduced an efficient algorithm to study trajectories within Gaussian random fields numerically. An interesting application for the future is to consider inflationary dynamics in high dimensional Gaussian random fields.

Turning towards the example of a simple Gaussian random supergravity, we studied the likelihood for metastable vacua. We found that at generic points where supersymmetry is badly broken by the F term, the probability for metastability is extremely suppressed and corresponds to a very unlikely matrix fluctuation. Based on heuristic arguments from random matrix theory we expect
\be
\log[P^{\text{generic}}_{\text{ensemble}}(\text{metastable c.p.})]\propto-N^2
\ee
at generic points. On the other hand, at points of approximate supersymmetry, where the supersymmetry breaking masses are small compared to the supersymmetry scale the probability of metastability is greatly enhanced and reduces to the study of the approximately supersymmetric regime in Ref.~\cite{Denef:2004cf,Marsh:2011aa} where
\be
\log[P^{\text{approx. SUSY}}_{\text{ensemble}}(\text{metastable c.p.})]\propto-N\,.
\ee
These points of approximate or exact supersymmetry occur at values of the potential that are low compared to generic points. Therefore, the vast majority of metastable vacua lie in dynamical attractor regions of approximate supersymmetry. It would be interesting to investigate the relative abundance of vacua with approximate and exact supersymmetry for a single choice of flux.

Considering the inflationary properties of a Gaussian supergravity landscape we find that $\epsilon\sim \eta\sim \M^2/\Lambda_h^2$ at generic points. While these parameters are small compared to the slow roll parameters in a Gaussian random landscape, where $\epsilon\sim N \M^2/\Lambda_h^2$, it turns out that a fluctuation to small slow roll parameters either requires horizontal correlation lengths on the order of the Planck scale or a large matrix fluctuation that is statistically extremely costly. Therefore, we conclude that small slow roll parameters are non-generic in the Gaussian random supergravity presented in this work.

To study random supergravities we chose a trivial K\"ahler potential and a Gaussian random superpotential. Generically, a non-trivial K\"ahler potential will enter both the statistical ensemble of the random superpotential via the two-point function and the effective potential and its derivatives via K\"ahler and geometric covariance. These additional contributions may well affect the results found in this work and will be considered in a future investigation.

We developed a set of tools that can be applied to the further study of more realistic effective random landscapes and potential consequences for multifield inflation. We found that the inflationary slow roll parameters are not necessarily large at generic, high points in the landscape where no metastable minima exist. This suggests an interesting structure of the landscape where high in the potential there are inflationary trajectories while metastable minima accumulate at very small potential values. This promising structure merits further study of more realistic supergravity models.

\section{Acknowledgments}
I would like to thank Mafalda Dias, Jonathan Frazer, Cody Long, David Marsh, Paul McGuirk, Enrico Pajer, John Stout and Timm Wrase for helpful discussions. I am particularly grateful to Liam McAllister for valuable discussions and comments on the draft. This work was supported by the NSF under grant PHY-0757868.

\bibliographystyle{modifiedJHEP}
\bibliography{refs}
\end{document}